\documentclass[a4paper,usenatbib]{mnras}
\usepackage{graphicx}
\usepackage[fleqn]{amsmath}
\usepackage[T1]{fontenc}
\usepackage{aecompl}
\usepackage{times}
\usepackage{ae,aecompl}
\usepackage{amssymb,amsfonts,hyperref,color}

\newcommand\be{\begin{equation}}
\newcommand\en{\end{equation}}

\title[Spirals]{A Simple Time-Dependent Method for Calculating Spirals: Applications to Eccentric Planets in Protoplanetary discs}
\author[Z.~Zhu \& R.M.~Zhang]{
Zhaohuan Zhu$^{1}$\thanks{E-mail: zhaohuan.zhu@unlv.edu},  Raymond M. Zhang$^{1,2}$\\
$^{1}$Department of Physics and Astronomy, University of Nevada, Las Vegas, 4505 S.~Maryland Parkway, Las Vegas, NV~89154, USA\\
$^{2}$Ed W. Clark High School, Las Vegas, NV~89154, USA\\}
\date{In original form \today}

\begin{document}
\label{firstpage}
\pagerange{\pageref{firstpage}--\pageref{lastpage}} 
\maketitle

\begin{abstract}
Spirals in protoplanetary discs have been used to locate the potential planet in discs.  Since only the spiral shape from a circularly orbiting perturber is known, most previous works assume that the planet is in a circular orbit. We develop a simple semi-analytical method to calculate the shape of the spirals launched by an eccentric planet. We assume that the planet emits wavelets during its orbit, and the wave fronts of these propagating wavelets form the spirals. The resulting spiral shape from this simple method agrees with numerical simulations exceptionally well. The spirals excited by an eccentric planet can detach from the planet, bifurcate, or even cross each other, which are all reproduced by this simple method. The spiral's bifurcation point corresponds to the wavelet that is emitted when the planet's radial speed reaches the disc's sound speed. Multiple spirals can be excited by an eccentric planet (more than 5 spirals when $e\gtrsim0.2$). The pitch angle and pattern speed are different between different spirals and can vary significantly across one spiral. { The spiral wakes launched by  high-mass eccentric planets steepen to spiral shocks and the crossing of spiral shocks leads to distorted or broken spirals. With the same mass, a more eccentric planet launches weaker spirals and induces a shallower gap over a long period of time.} The observed unusually large/small pitch angles of some spirals, the irregular multiple spirals, and the different pattern speeds between different spirals may suggest the existence of eccentric perturbers in protoplanetary discs. 

\end{abstract}


\section{Introduction}

Spirals are ubiquitous in astrophysical disc systems. 
On large scales, spiral galaxies show  spiral structures extending from the center to the edge of galactic discs \citep{Binney1998}. 
The spiral arms are sites of star formation, and some spirals can be observed from X-ray all the way to radio wavelengths. On small scales,
spirals are found in Saturn's rings \citep{Cuzzi1981}. Spirals with various $m$ symmetry are found in {\it Cassini} images. { Spirals have been discovered
in protoplanetary discs thanks to advancements in adaptive optics and radio interferometry.} Some discs show grand-design two spirals in near-IR polarized images \citep{Muto2012,Stolker2017,Benisty2015,Wagner2015} and submm dust continuum images \citep{Perez2016,Huang2018b}, while some discs show
multiple spiral arms \citep{Avenhaus2014,Follette2017,Monnier2019,Boccaletti2020}.

The universal spiral structure in discs is explained by the density wave theory \citep{Binney2008,Shu2016}. 
Sound waves (density waves) that propagate
in the disc adopt spiral forms (details in Section 2). These density waves can be excited by perturbers \citep{GoldreichTremaine1979}, turbulence \citep{Heinemann2009}, disc self-gravity \citep{Bethune2021,Baehr2021}, vortices \citep{Paardekooper2010}, or the central object's gravitational potential (e.g., bared galaxies and Saturn's ring seismology \citealt{Mankovich2019}). Independent from the excitation mechanism, the waves will propagate freely after the excitation. 

When the amplitude of a free propagating wave is small, the shape of the spiral is mainly determined by the disc properties (e.g., disc temperature) and
the wave's dispersion relationship. The single-arm spiral excited by a low-mass perturber in a circular orbit is studied by \cite{Ogilvie2002}. This spiral
 is the result of constructive interference between various $m$ modes. 
These modes have small dispersions among them, which leads to the formation of multiple spirals when the waves propagate far away from the planet
 \citep{Bae2018,Miranda2019}. When the spiral's amplitude is large (e.g., excited by a massive planet), the non-linear wave steepening effect becomes important \citep{Goodman2001}, and  a spiral wake becomes a spiral shock. The stronger the shock is, the larger the spiral's opening angle is. The multiple spirals excited by a high-mass planet are well separated, forming the $m=2$ grand-design structure in the disc \citep{Zhu2015c, Fung2015, Bae2018b}. 

While the theoretical works with a circularly orbiting perturber have some success in explaining spirals in protoplanetary disc observations \citep{Dong2015, Dong2016,Dong2017,Rosotti2020,Baruteau2019}, there are notable discrepancies between theory and observations.  First, the planets predicted in some spiral systems have not been discovered/confirmed (e.g., \citealt{Boccaletti2021}). Second, the pitch angles of observed spirals vary greatly from less than 5$^o$ \citep{Teague2019} to 30$^o$ \citep{Monnier2019}, which are both difficult to be explained unless the planet is very far or very close to the spirals. Third, many systems show multiple spiral arms \citep{Avenhaus2014, Monnier2019,Boccaletti2020}, while the planet-induced spirals normally show 1 or 2 spirals. Fourth, the two spiral arms in SAO 206462 may have different pattern speeds \citep{Xie2021}, which cannot be explained by a circularly moving perturber with a constant pattern speed.
Although these discrepancies can be due to the projection effect for inclined systems \citep{Follette2017} or a different spiral excitation mechanism \citep{Dong2015b,Bae2021}, they may also be due to that the planet is in an eccentric orbit. 
\cite{Calcino2020} use an eccentric perturber to reproduce several features of the spirals in MWC 758. The eccentric perturber has also been proposed to explain observed gap structures \citep{Muley2019,Li2019,Chen2021}. Theoretical works also suggest that planets can gain eccentricity by interacting with protoplanetary discs \citep{Goldreich2003,Teyssandier2017,Ragusa2018}. { For a luminous pebble accreting planet, the planet's eccentricity can be larger than the disc's aspect ratio \citep{Velasco-Romero2021}. } Recent simulations by \cite{Baruteau2021} suggest that the orbit of a giant planet can reach $e=0.25$ when the planet migrates into a cavity.

In this paper, by drawing an analogy to Huygens' principle, we develop a time-dependent method to calculate the shape of the spirals launched by an eccentric planet in a disc. The method is introduced in Section 2, and it is compared against numerical simulations in Section 3. After studying the spirals' pitch angles and pattern speeds in Section 4, we conclude the paper in Section 5.

\section{The Analytical Method}

\subsection{Background}
\begin{figure*}
\includegraphics[trim=30mm 20mm 20mm 10mm, clip, width=3.3in]{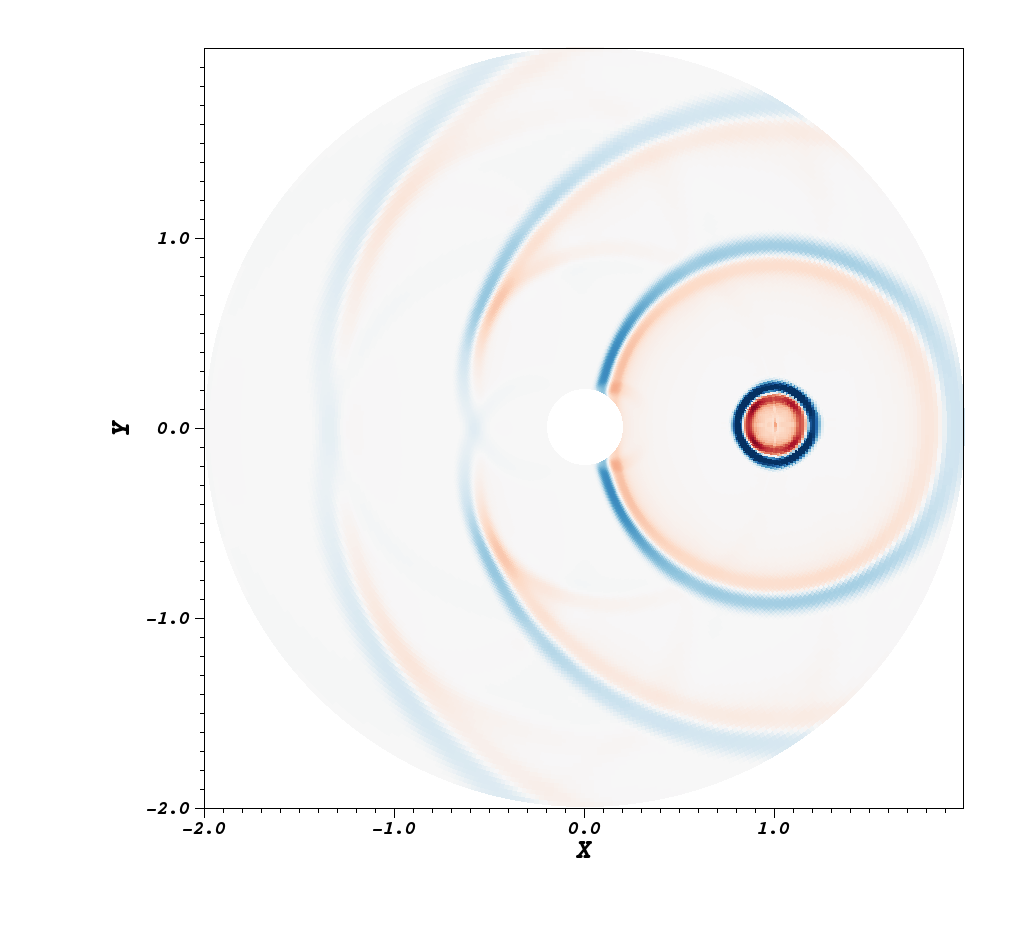}
\includegraphics[trim=30mm 20mm 20mm 10mm, clip, width=3.3in]{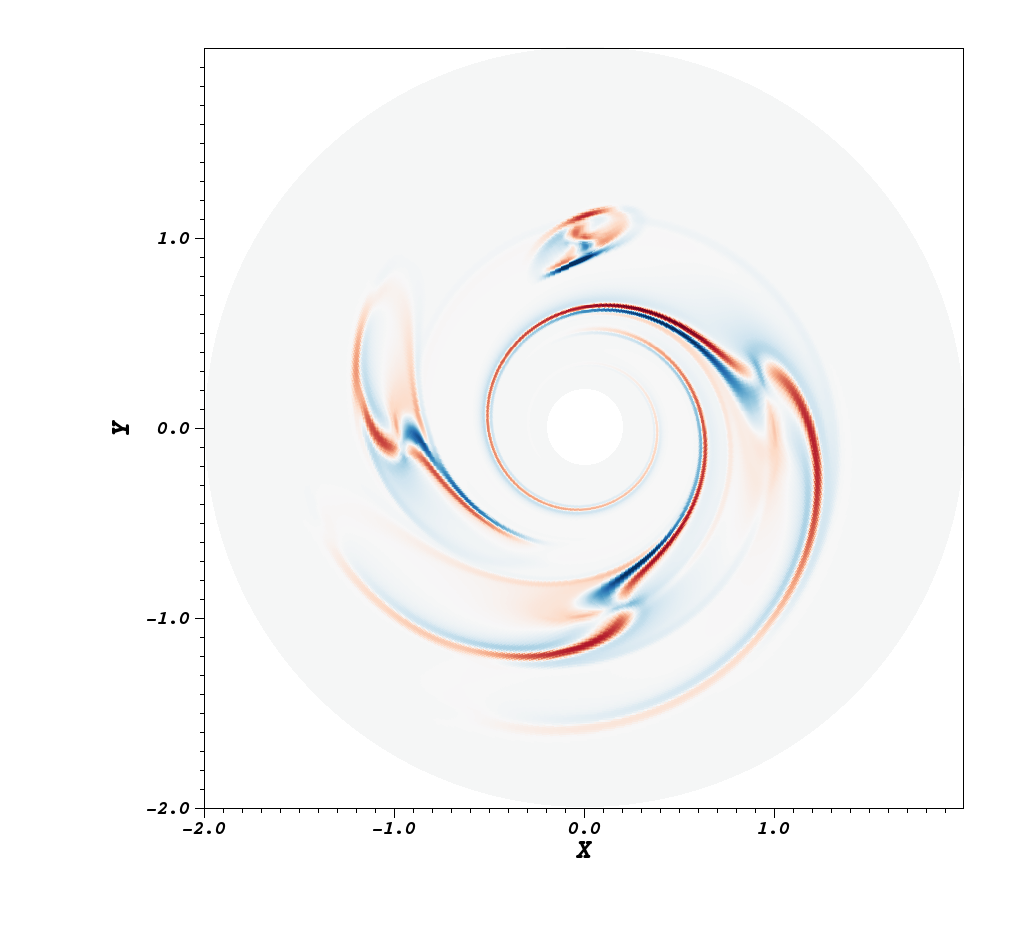}\\
\caption{Density disturbance in a non-rotating system (the left image) and a Keplerian rotating system (the right image) 
after the fluid has been disturbed four times at X=1, Y=0. The simulations are carried out using Athena++ under a $R-\theta$ 2-D polar coordinate system. 
Both density and sound speed are constant throughout the whole region. }
\label{fig:visit}
\end{figure*}
Perturbation in a medium can lead to visible disturbances or patterns which propagate in the medium. 
We use wakes to refer to these visible disturbances or patterns, while we use waves to refer to individual
propagating modes which satisfy the wave dispersion relationship. The wakes form from the interference of all individual 
wave modes. If all individual wave modes lead to positive enhancements at the same position, the final disturbance 
will be positive. If there are a mixture of positive and negative contributions from different wave modes at the same position, these waves
may cancel out each other and lead to little final disturbance.

The left panel of Figure 1 shows the wake pattern after we perturb a static uniform medium at the same position 
 for four times. Specifically, we increase the density at x=1, y=0 instantaneously and then reset it back to the initial density. After
each perturbation, a ring of disturbance is launched which is propagating spherically outwards at the sound speed. The disturbance
gets weaker while it propagates further. The right panel of Figure 1 shows the wave pattern after we perturb a Keplerian disc
at the same position in the inertial frame (x=1, y=0) for four times. After each perturbation, the disturbance is quickly advected azimuthally away from the
perturbation point (x=1, y=0) following the local Keplerian flow. At the same time, the disturbance launches a spiral which corotates with
the disturbance with the pattern speed $\Omega_{pat}=\Omega_{K}(R=1)$. The spiral propagates further inwards and outwards with time.
Clearly, a disturbance propagates differently in a rotating disc than in a static medium. A disturbance in a shearing disc is 
always in a spiral form, which leads to the ubiquitous spirals found in disc observations of spiral galaxies, protoplanetary discs, and Saturn's rings.

\begin{figure}
\includegraphics[trim=10mm 50mm 50mm 10mm, clip, width=3.3in]{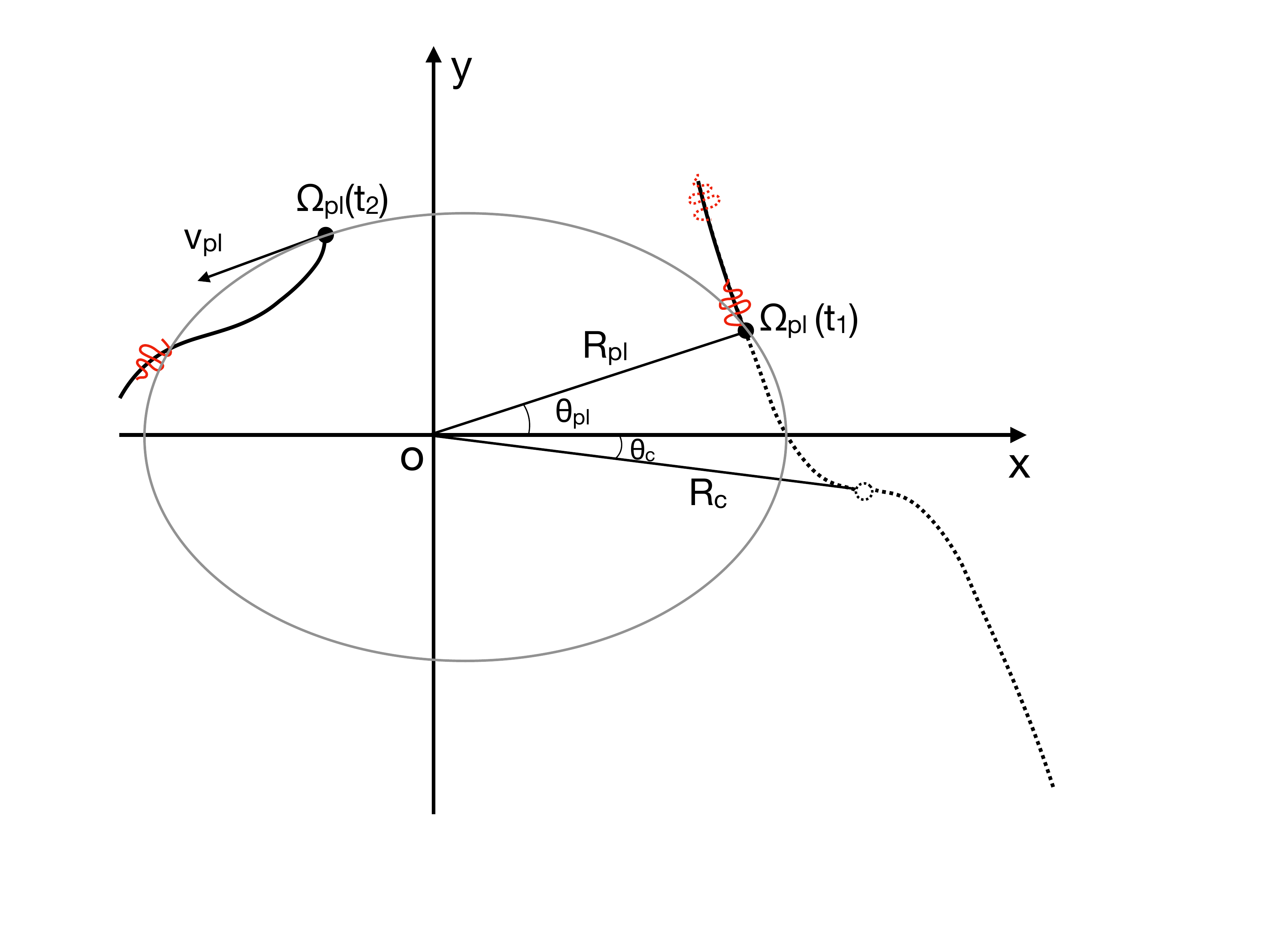}\\
\caption{ The spirals at $t=t_2$ are the sum of all the wavelets that the planet emitted in the past during the planet's orbital motion. 
The wavelet of the spiral at $t_2$ (the red wiggled curve) is excited by the planet at $t_1$. The wavelet propagates following the spiral shape while rotates
with the planet's orbital frequency at $t_1$ (labeled as $\Omega_{pl}(t_1)$, which is the wavelet's pattern speed).  The spiral shape can be derived analytically using the planet's corotating radius $R_c$.}
\label{fig:ana}
\end{figure}
 
If the source of the perturbation is moving (e.g., a traveling boat in the water or an orbiting planet in a protoplanetary disc), the wake will have a different pattern
since different sets of waves can interfere coherently. In a static medium, a Mach cone forms after an object that moves at the supersonic speed of v. The 
opening angle of the Mach cone is arcsin($c_{s}/v$) since only waves with  the wavevector ${ k}\cdot { v}=c_{s}$ maintain the same phase 
with respect to the moving object so that a coherent interference can occur.  { For a ship moving in water, the dispersion of the water waves leads to the Kevin wake pattern.} For a planet orbiting around the star at the local Keplerian speed, 
all the excited spiral
waves interfere with each other coherently far away from the planet to form a single-arm spiral, since spiral waves with different $m$ modes have the same pitch angle far away from the perturber \citep{Ogilvie2002}. Specifically, considering that the WKB dispersion relationship for spiral waves \citep{Binney2008}  in a non-self-gravitating disc is
\begin{equation}
m^2(\Omega_{pat}-\Omega)^2=\kappa^2+c_{s}^2 k^2\,,
\label{eq:disper}
\end{equation}
where { $k$ is the wavenumber } and $\kappa$ is the epicycle frequency which equals $\Omega_{k}$ in a Keplerian disc,  the pitch angles ($\zeta$) of all the modes ($\zeta$=arccot $|kR/m|$) are approaching arccot$|(\Omega_{pat}-\Omega)R/c_s|$ when they are far away from the planet so that $\kappa^2$ can be ignored. With the same pitch angle, the different $m$-mode waves can interfere with each other coherently, forming a single-arm spiral. If we integrate the pitch angle starting from the planet's position, we can derive that the spiral arm follows
\begin{eqnarray}
\theta&=&\theta_{pl} + sgn(R-R_{pl})\frac{R_{pl}\Omega(R_{pl})}{c_{s}(R_{pl})}\nonumber\\
&&\Bigg[-\left(\frac{R}{R_{pl}}\right)^{\beta+1}\left(\frac{1}{1+\beta}-\frac{1}{1-\alpha+\beta}\left(\frac{R}{R_{pl}}\right)^{-\alpha}\right)\nonumber\\
&&+\left(\frac{1}{1+\beta}-\frac{1}{1-\alpha+\beta}\right)\Bigg]\,,\label{eq:anaint}
\end{eqnarray}
in a disc with $\Omega(R)\propto R^{-\alpha}$ and the sound speed of $c_{s}(R)\propto R^{-\beta}$ \citep{Rafikov2002,Muto2012}.
From Equation \ref{eq:disper}, we can see that forming a single spiral arm is a property of the wave propagation and independent from the origin of the perturbations (whether due to the presence of a planet or some other density fluctuations). 

To study wakes excited by a circularly orbiting planet,  the rotating frame is normally adopted so that a steady state can be achieved. This significantly simplifies the analysis and a time independent solution can be found \citep{Ogilvie2002,Miranda2019}. However, it is difficult to extend this approach to time-dependent phenomena, e.g., perturbation by a planet in an eccentric orbit. 

\subsection{The Time-dependent Method}
By drawing an analogy to Huygens' principle, we adopt a time-dependent approach to study spirals launched by an eccentric perturber in a disc. Huygens' principle states that every point of a wave front can be regarded as new sources of wavelets. After these wavelets propagate for a while in the spherical fashion, the surface which is tangent to the spherical wavelets, called the envelope of the wavelets, is the new wave front.
In other words,  current disturbance at any point is influenced by all the points in the past that can propagate to this current point. Such a principle also applies to fluid dynamics. Furthermore, thanks to the fluid equations' linearity for small perturbations, we can study how the small disturbance at every position in the past contributes to the current disturbance, and linearly add together the disturbances from all these positions in the past to derive the current wake. 

This allows us to separate all disturbances at one time (including both the existing spiral and the perturbation around the planet) into small wavelets, then follow the wavelets' propagation, and finally add all the wavelets at a later time to study how the wake changes with time. We can apply this approach to study the shape of spirals excited by an eccentric perturber in a disc, but with several simplifications to make the problem trackable. First, we assume that every piece of the spiral (wavelet) propagates in the radial direction at the local sound speed. A wavelet in a non-self-gravitating disc propagates at the group velocity of 
\begin{equation}
v_{g}(R)=sgn(k)\frac{k c_{s}^2}{m(\Omega_{pat}-\Omega)}\,.\label{eq:vgr}
\end{equation}
Plugging in Equation \ref{eq:disper}, we can see $v_{g}(R)\approx sgn(k)c_{s}$ when $\kappa$ is dropped, e.g., far away from the position of wave excitation. 
Thus, trailing wavelets ($k>0$) propagate outwards while leading wavelets ($k<0$) propagate inwards.  Second, we assume that every wavelet follows
the pitch angle of arccot$|(\Omega_{pat}-\Omega)R/c_s|$ while it propagates either inwards or outwards and rotates at the pattern speed $\Omega_{pat}$. Note that $\Omega_{pat}$ is the azimuthal pattern speed of this particular wavelet and different parts of the spiral can have different $\Omega_{pat}$. Third, we assume that every wavelet maintains its $\Omega_{pat}$ during its propagation. The second and third assumptions are based on the non-dispersive properties of different $m$-modes far away from the perturber (Equation \ref{eq:disper}). All $m$-modes maintain their interference with a constant $\Omega_{pat}$ so that they follow the spiral shape during propagation. These assumptions are also justified in Figure 1, where any perturbation propagates inwards/outwards and follows the spiral shape of Equation \ref{eq:anaint}. If we look back in time,  every current disturbance can eventually be traced back to the perturbation at the planet's location when the wavelet was just launched. The disturbance's $\Omega_{pat}$ equals the planet's angular frequency back then. 

Thus, we can think that the planet is emitting wavelets all the time during its orbital motion, and a wavelet propagates at the sound speed in the radial direction while following a spiral shape in the azimuthal direction. As long as we know the planet's position with time, we can determine the wavelets' positions at any time and we can determine the spiral shape by simply connecting all the wavelets. Imagine that two wavelets are launched by a planet at time $t_1$ (one traveling inwards and one traveling outwards), the wavelets thus have a pattern speed of $\Omega_{pat}=\Omega_{pl}(t_1)$, where $\Omega_{pl}(t_1)$ is the planet's angular frequency at time $t_1$. At time $t_2$, these wavelets propagate to the radial location at
\begin{equation}
R_{t_1}(t_2)=R_{pl}(t_1)-\int_{t_1}^{t_2}c_s dt \label{eq:rinner}
\end{equation} 
for the inner spiral and 
\begin{equation}
R_{t_1}(t_2)=R_{pl}(t_1)+\int_{t_1}^{t_2}c_s dt \label{eq:router}
\end{equation}
for the outer spiral, where $R_{pl}(t_1)$ is the planet's radial location at time $t_1$.  Assuming $c_{s}(R)\propto R^{-\beta}$, we can derive
\begin{equation}
R_{t_1}(t_2)^{\beta+1}-R_{pl}(t_1)^{\beta+1}=\pm\left(\beta+1\right)c_{s}(R_{pl}(t_1))R_{pl}(t_1)^{\beta}\left(t_2-t_1\right)\,.
\end{equation}
To calculate the azimuthal locations of these wavelets at $t_2$, we can integrate
\begin{equation}
 cot\zeta\equiv Rd\theta/dR=|(\Omega_{pat}-\Omega)R/c_s|\label{eq:pitchangle}
\end{equation} 
starting from the planet's radial position at $t_1$ to $R_{t_1}(t_2)$ (Equations \ref{eq:rinner},\ref{eq:router}), where $\Omega_{pat}=\Omega_{pl}(t_1)$ being a constant. After deriving $\theta$, we need to add an additional $\Omega_{pat}(t_2-t_1)$ to $\theta$ to account for the pattern's rotation in the inertial frame. On the other hand, Equation \ref{eq:anaint} is derived from integrating $Rd\theta/dR$. Thus, we can use Equation \ref{eq:anaint} to calculate $\theta$ of the wavelet at $t_2$ analytically. First, at $t_1$, we need to find the corotation radius
where $\Omega(R_c)=\Omega_{pl}(t_1)$ (Figure \ref{fig:ana}). Then, the wavelet can be considered as traveling along the spiral of the imaginary planet at ($R_c$, $\theta_{c}$). $\theta_c$ can be derived using Equation \ref{eq:anaint} by replacing $\theta$ with $\theta_{pl}(t_1)$, $\theta_{pl}$ with $\theta_c$, $R$ with $R_{pl}(t_1)$, and $R_{pl}$ with $R_c$. Eventually, we can derive
\begin{eqnarray}
\theta_c=&\theta_{pl}(t_1)-sgn(R_{pl}(t_1)-R_{c})\frac{R_{c}\Omega(R_{c})}{c_{s}(R_{c})}\nonumber\\
&\Bigg[-\left(\frac{R_{pl}(t_1)}{R_{c}}\right)^{\beta+1}\left(\frac{1}{1+\beta}-\frac{1}{1-\alpha+\beta}\left(\frac{R_{pl}(t_1)}{R_{c}}\right)^{-\alpha}\right)\nonumber\\
&+\left(\frac{1}{1+\beta}-\frac{1}{1-\alpha+\beta}\right)\Bigg]\,,
\end{eqnarray}
where $\theta_{pl}(t_1)$ is the planet's azimuthal position at $t_1$.
Then, with
$R_c$ and $\theta_c$ derived, we can find $\theta$ of the wavelet at $t_2$ by replacing $R$ in Equation \ref{eq:anaint} with Equations \ref{eq:rinner} and \ref{eq:router}, $R_{pl}$ with  $R_c$, and $\theta_{pl}$ with $\Omega_{pl}(t_1)(t_2-t_1)+\theta_c$ which is the imaginary planet's position at $t_2$. We thus have
\begin{eqnarray}
\theta=&\Omega_{pl}(t_1)(t_2-t_1)+\theta_c+ sgn(R_{t_1}(t_2)-R_{c})\frac{R_{c}\Omega(R_{c})}{c_{s}(R_{c})}\nonumber\\
&\Bigg[-\left(\frac{R_{t_1}(t_2)}{R_{c}}\right)^{\beta+1}\left(\frac{1}{1+\beta}-\frac{1}{1-\alpha+\beta}\left(\frac{R_{t_1}(t_2)}{R_{c}}\right)^{-\alpha}\right)\nonumber\\
&+\left(\frac{1}{1+\beta}-\frac{1}{1-\alpha+\beta}\right)\Bigg]\,.
\end{eqnarray}
Through these steps, 
we find the position of the wavelet
at $t_2$ that is emitted by the planet at $t_1$. After dividing the planet's orbital motion into many wavelet emitting events separated with a time interval of $\Delta T$, we can calculate 
the positions of all wavelets at $t_2$ analytically. By connecting all the wavelets at $t_2$ that are emitted by the planet from t=0 to t=$t_2$ with the interval of $\Delta T$, we can derive the shape of the spiral at $t_2$.  Although this method is not fully analytical at the final step, we will still call it the analytic method to be compared with numerical simulations later.

\section{Numerical Simulations and Comparison}

To verify the analytical method, we use FARGO code \citep{Masset2000} to carry out 2-D hydrodynamical simulations to study wakes excited by planets.
We have carried out 4 simulations with different planet eccentricities (e=0, 0.1, 0.25, 0.5). The planet mass is very low with $M_p/M_{*}=3\times 10^{-6}$ (equivalent to $M_p=M_{\oplus}$  if $M_*=M_{\odot}$) so that the excited wakes are in the linear regime. { In \S 5.2, we have increased the planet mass to one Jupiter mass to study the spirals excited by high-mass planets.} The planet is on fixed orbits with a gravitational smoothing length of 0.3 disc scale height. { Indirect forces have not been included in the low planet mass cases while they have been included in the high planet mass cases.} The radial domain extends from 0.1 to 5 $R_p$ with 638 radial grids that are uniformly spaced in log$R$ and 1024 azimuthal grids that are uniformly spaced in $\theta$. 
The disc's aspect ratio ($c_{s}/v_{K}$) is 0.1 at $R=R_p$. The disc has a uniform surface density and a constant temperature to simplify the analysis. Thus, $\alpha=1.5$ and $\beta=0$. We have only run the simulations for 4 planetary orbits to study how the spirals are launched. The sound crossing time is $R/c_{s}$ which is around 1.6 orbits. The wakes reach steady states after 2 orbits. To compare with these numerical simulations, our analytical study adopts a time interval of $\Delta T$=1/200 $T_{orb}$ between every wavelet emitting event by the planet. 

\begin{figure}
\includegraphics[trim=10mm 0mm 5mm 10mm, clip, width=3.3in]{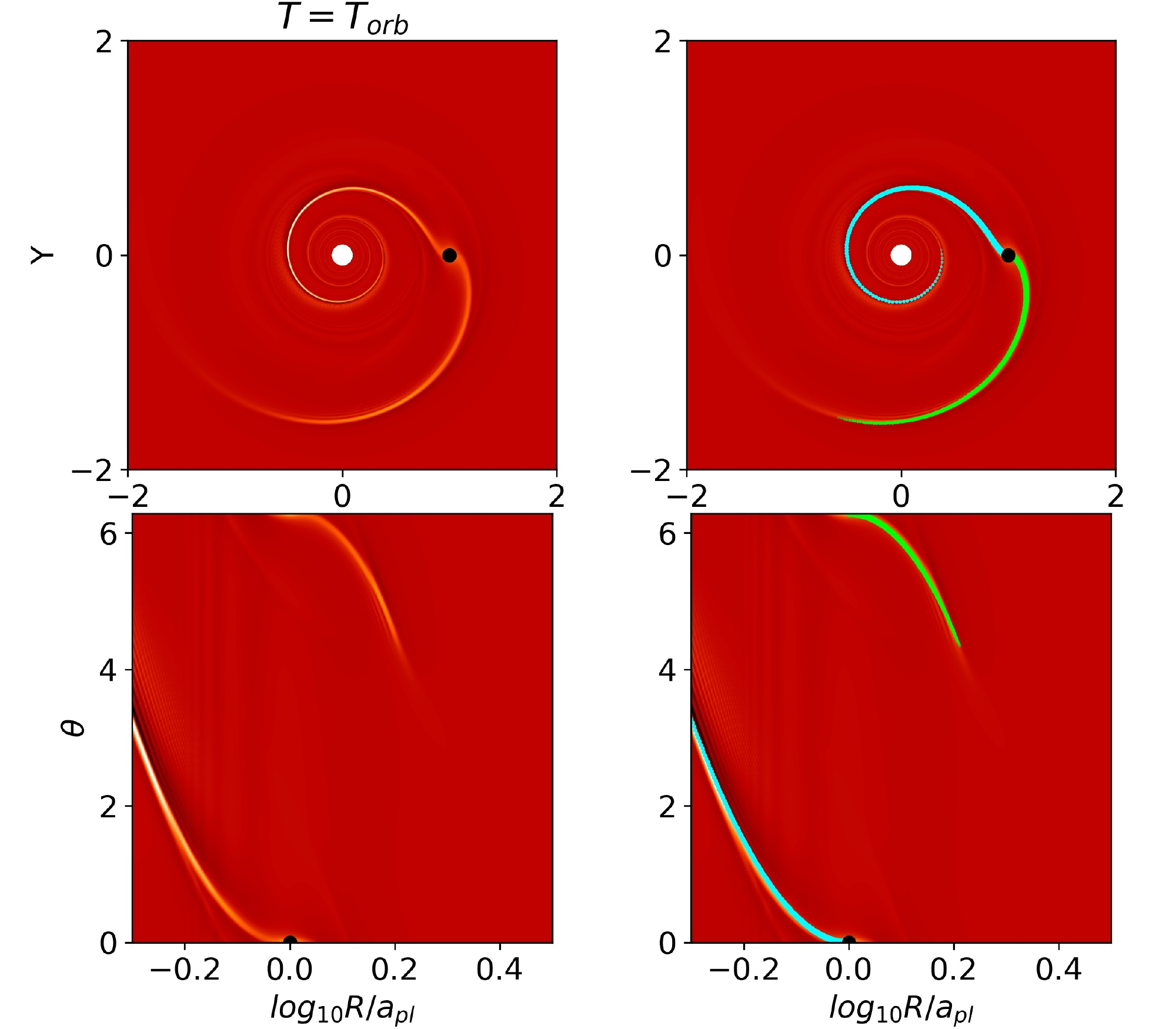}\\
\caption{The spirals excited by a circularly orbiting planet at t=$T_{orb}$. The spirals are shown in the Cartesian coordinate system (top panels) and the polar coordinate system (bottom panels). The right panels overplot the shape of spirals from the analytical theory (cyan curves: the inner spirals, green curves: the outer spirals).   }
\label{fig:circular}
\end{figure}

\begin{figure}
\includegraphics[trim=50mm 20mm 50mm 20mm, clip, width=3.3in]{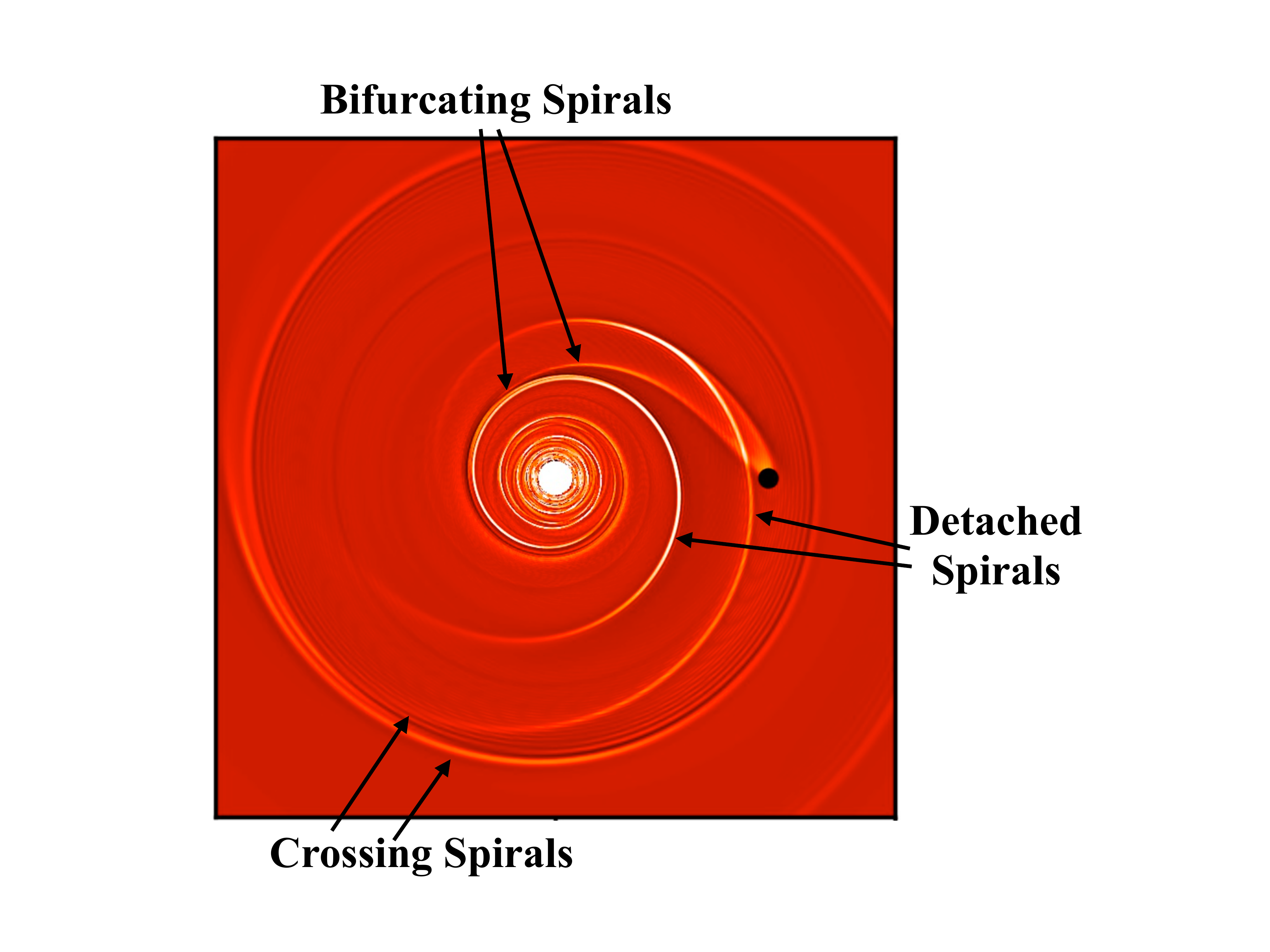}\\
\caption{The spirals excited by an eccentric planet with $e=0.25$ at 2$T_{orb}$. We can see spirals that are detached from the planet and detached from each other, spirals that bifurcate, and spirals that cross each other. The movies can be downloaded at \url{https://www.physics.unlv.edu/~zhzhu/Movies10.html} }
\label{fig:untitled3}
\end{figure}

The wakes at 1 planetary orbit ($T_{orb}$) with the circular perturber are shown in Figure \ref{fig:circular} (top rows:
in the Cartesian coordinate system; bottom rows: in the polar coordinate system).  Both the inner and outer wakes have not yet reached
to the inner and outer boundaries. The spirals from the analytical theory are traced by cyan and green dots in the right panels. The dots represent
wavelets that are emitted at a time interval of  1/200 $T_{orb}$. The earlier the wavelet is admitted, the smaller the dot is. 
The dots are so packed together (especially close to the planet) that they form a continuous curve.
The cyan dots trace the inner spiral which uses
$R_{t_1}$ from Equation \ref{eq:rinner}, while the green dots trace the outer spiral with Equation \ref{eq:router}.  As expected for the circular perturber, the analytical spirals follow Equation \ref{eq:anaint} and trace the spirals in simulations very well. Furthermore, the analytical spirals 
roughly show the radial extend of the spirals at $t=T_{orb}$. At the tips of the spiral arms, we can see that the simulated spirals extend a little bit further away than the analytical model. This is because the waves have intrinsic dispersion, and the wavelet is not localized as a single point as we assumed. We will discuss this caveat in Section 5.

\begin{figure*}
\includegraphics[trim=0mm 3mm 0mm 0mm, clip, width=6.8in]{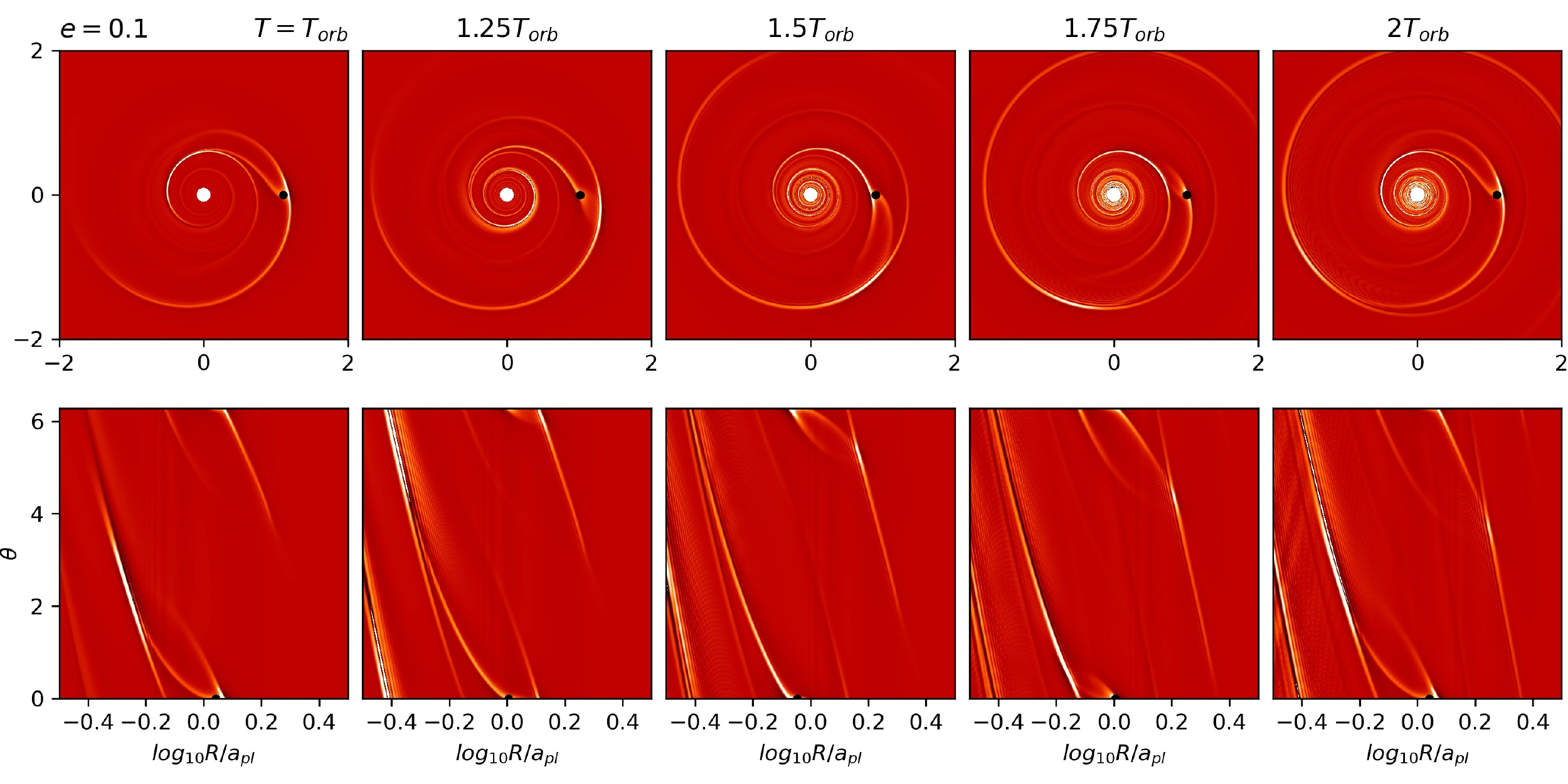}\\
\includegraphics[trim=0mm 3mm 0mm 0mm, clip, width=6.8in]{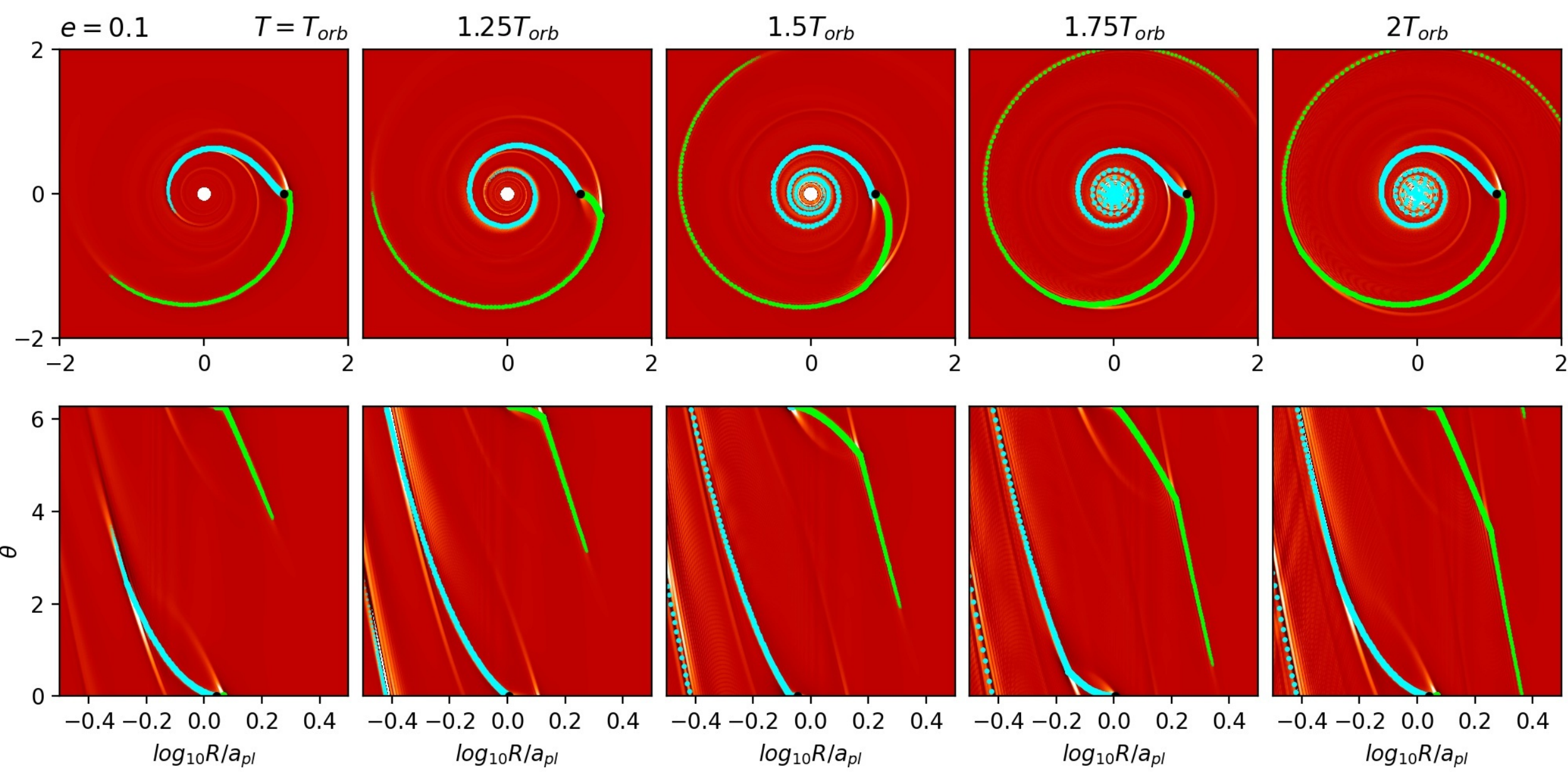}\\
\caption{Top two rows: the spirals excited by an eccentric planet with $e=0.1$ at different times from $T_{orb}$ to 2 $T_{orb}$ (from the left to right panels). The corotating frame with the planet is adopted, so that the planet is at the same azimuthal angle in all the plots. 
The spirals are shown in the Cartesian coordinate system (the first row) and the polar coordinate system (the second row).
Bottom two rows: similar to the top rows but with the spirals from the analytical theory overplotted. }
\label{fig:e0p1}
\end{figure*}

\begin{figure*}
\includegraphics[trim=0mm 3mm 0mm 0mm, clip, width=6.8in]{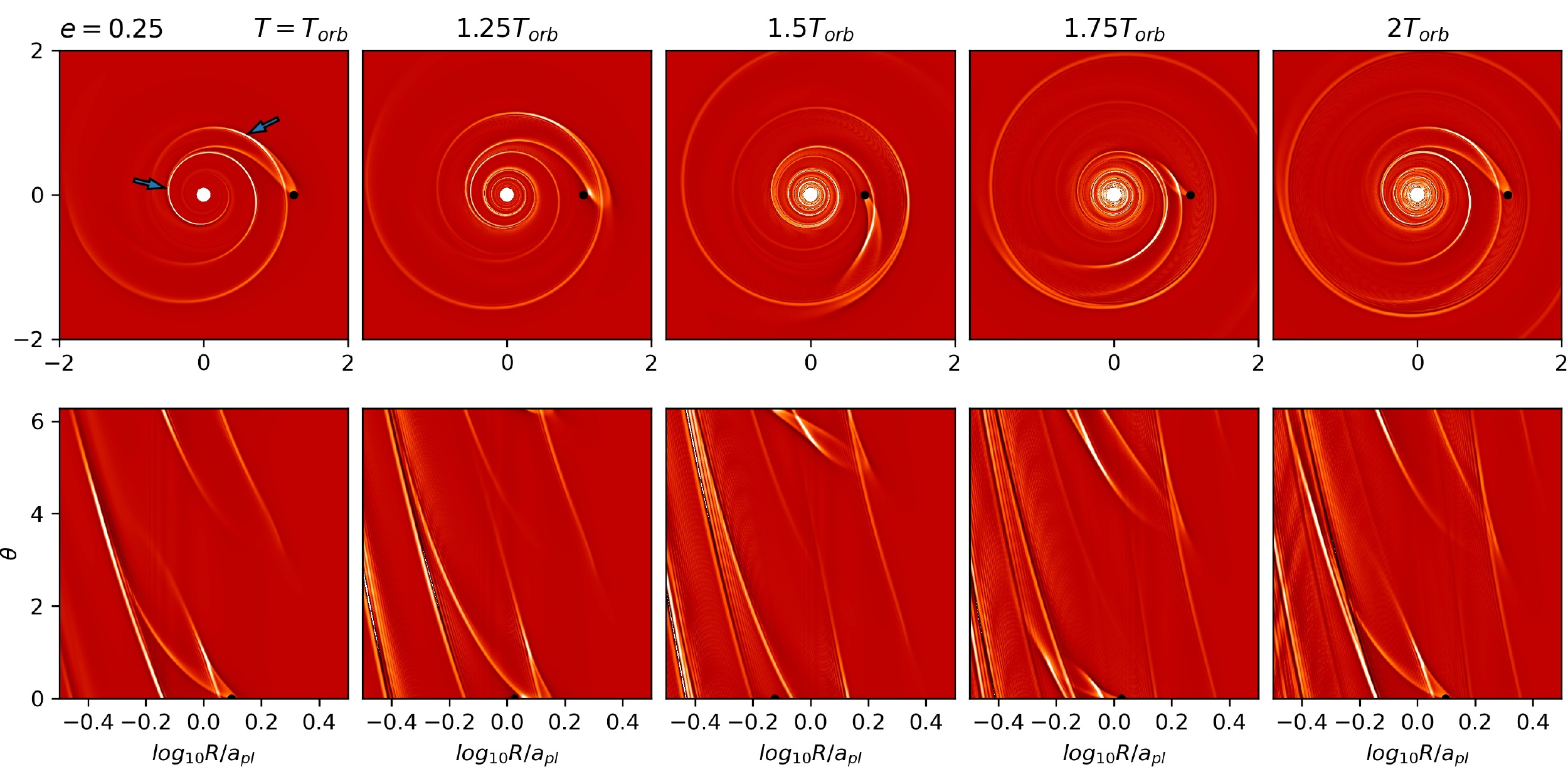}\\
\includegraphics[trim=0mm 3mm 0mm 0mm, clip, width=6.8in]{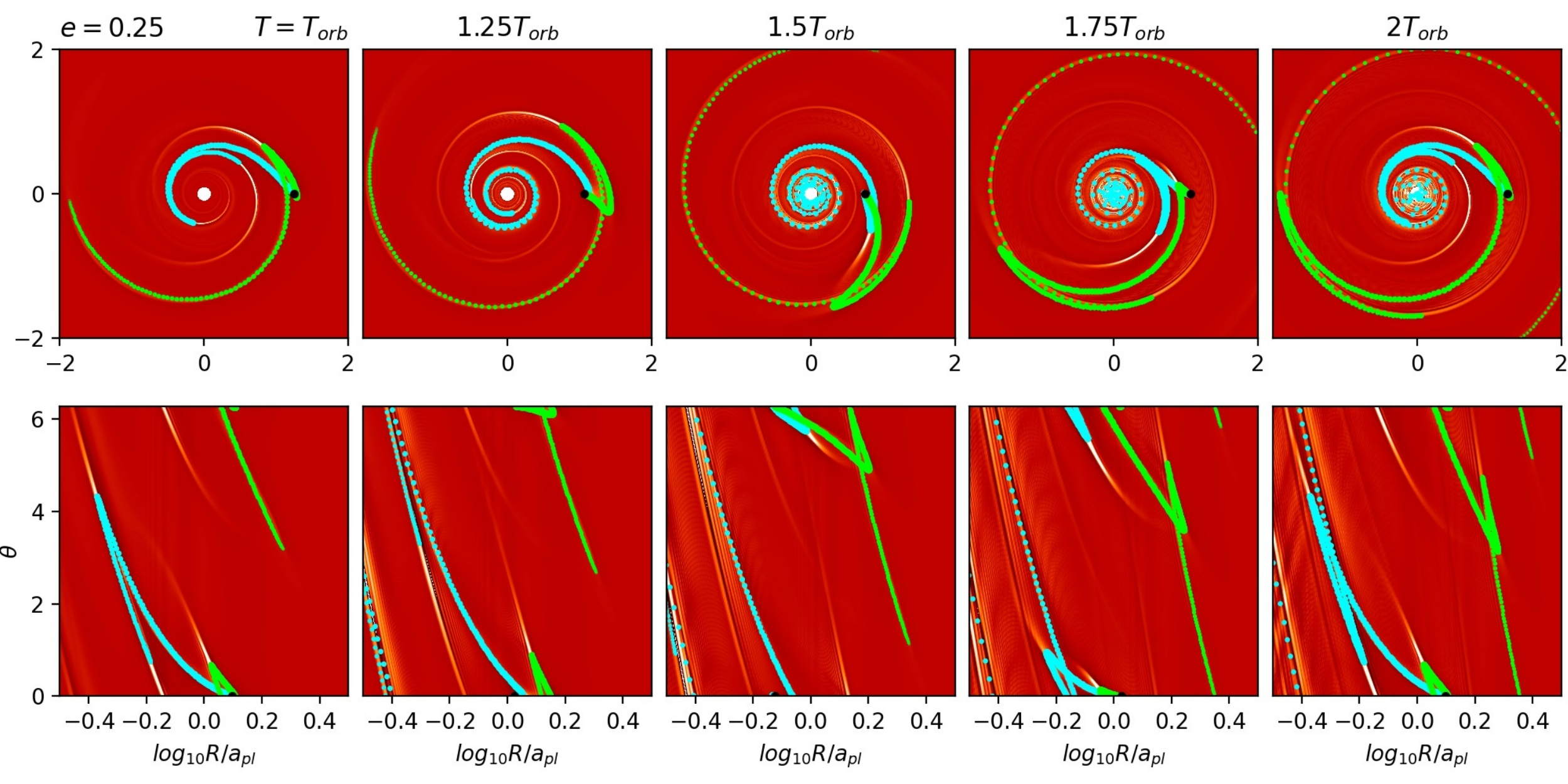}\\
\caption{ Similar to Figure \ref{fig:e0p1} but with an eccentric planet having $e$=0.25. The ``V points'' where the spirals bifurcate are labeled with the arrows.  }
\label{fig:e0p25}
\end{figure*}

\begin{figure*}
\includegraphics[trim=0mm 3mm 0mm 0mm, clip, width=6.8in]{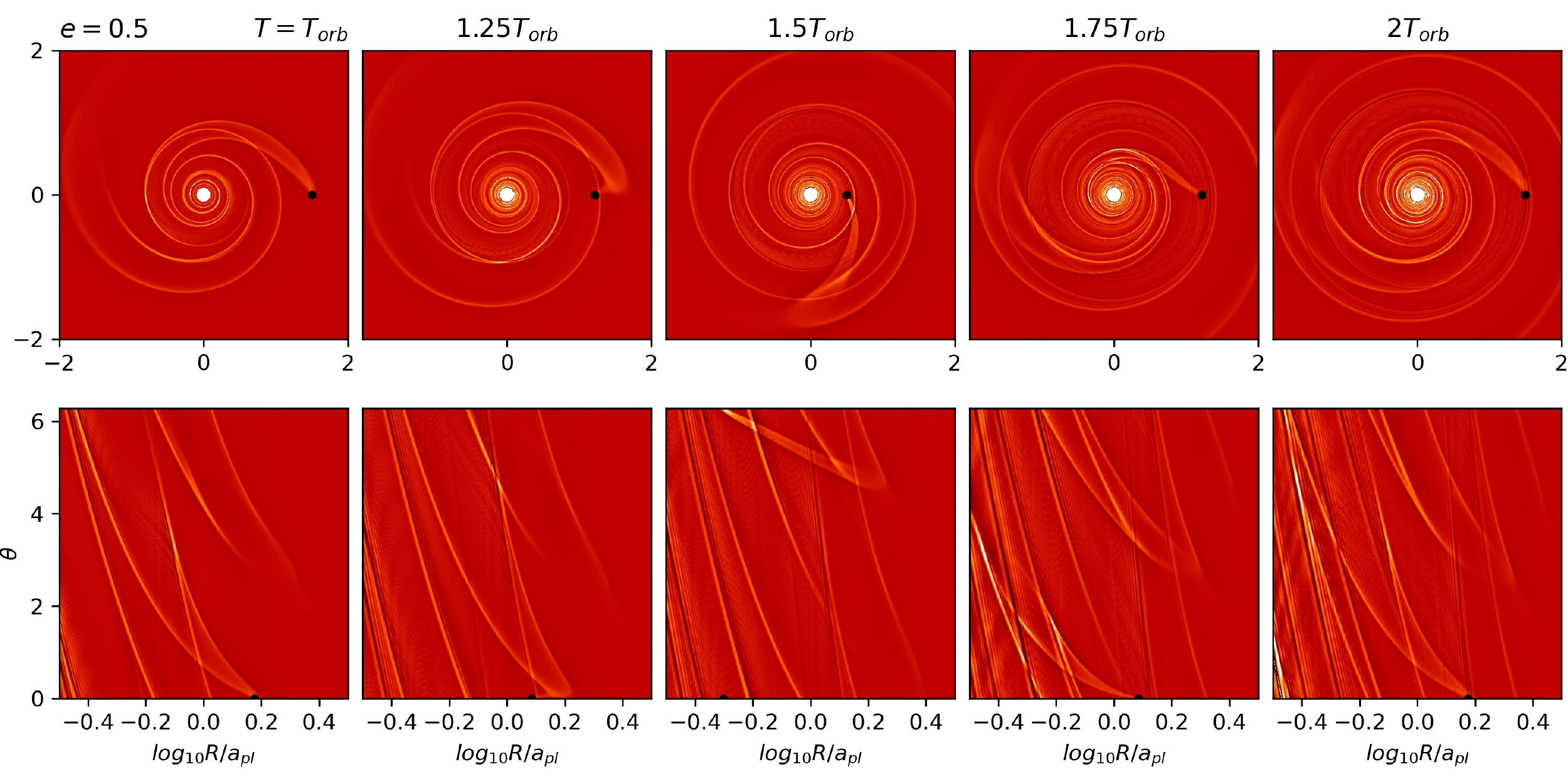}\\
\includegraphics[trim=0mm 3mm 0mm 0mm, clip, width=6.8in]{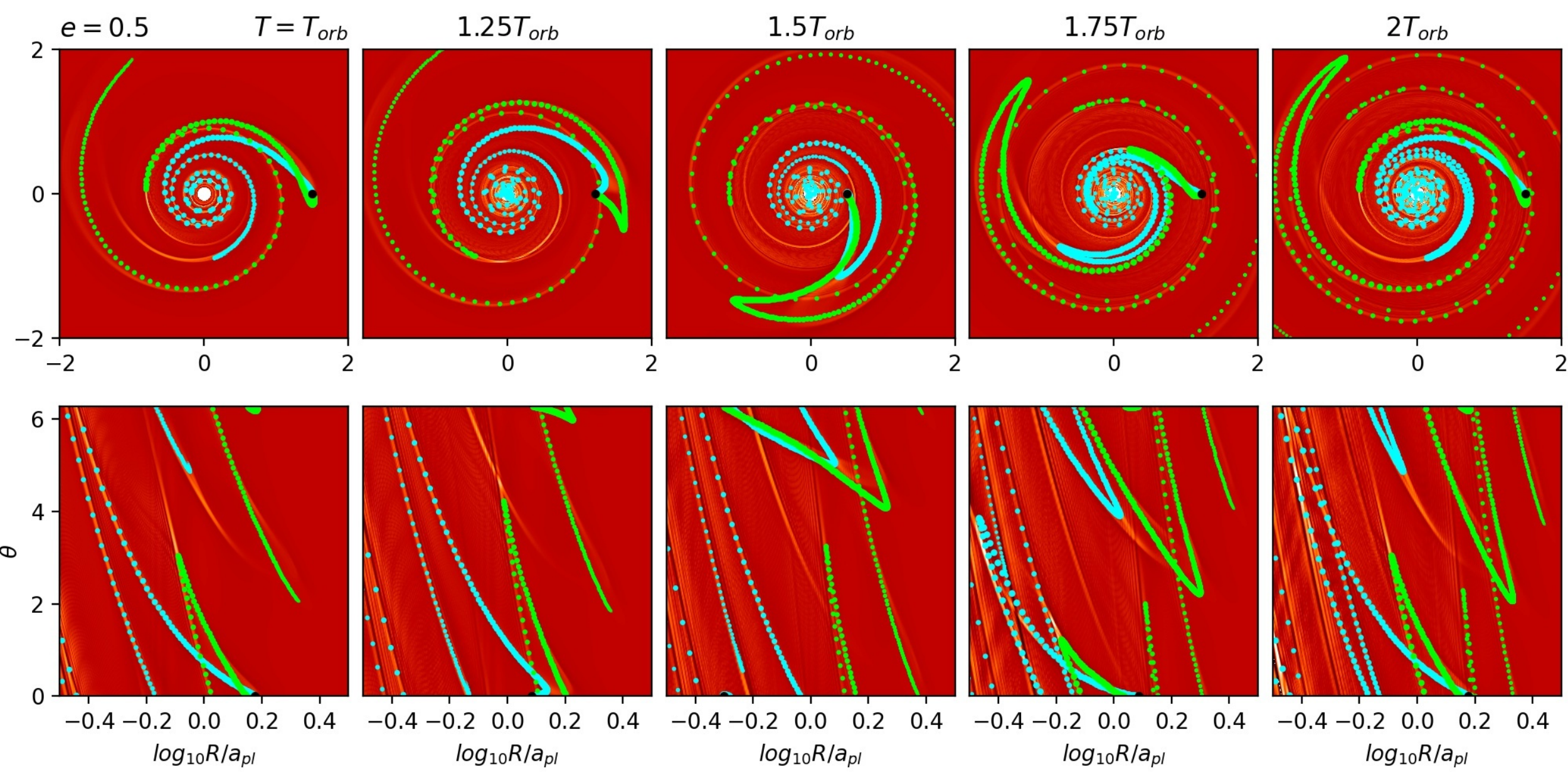}\\
\caption{ Similar to Figure \ref{fig:e0p1} but with an eccentric planet having $e$=0.5.}
\label{fig:e0p5}
\end{figure*}

The spirals excited by an eccentric perturber are quite different from the spirals by a circular perturber. 
With the perturber's $e$=0.25 (Figure \ref{fig:untitled3}), the excited spirals can detach from the planet.
Both the inner and outer spirals can exist inside the planet's position and they are separated from each other. { We define the inner/outer spirals as the wakes that propagate inwards/outwards from the planet after they are launched. It is difficult to distinguish them just from the images without the help of the analytical theory (the inner and outer spirals are labeled with different colors in Figures \ref{fig:e0p1} to \ref{fig:e0p5}).}
The spirals can bifurcate and even cross each other at some positions. Furthermore, the spiral's shape is changing
with time, as shown in Figures \ref{fig:e0p1} to \ref{fig:e0p5}.
We adopt the corotating frame where the planet is always at $y=0$, which highlights the spirals' movements with respect to the planet. 
In Figure \ref{fig:e0p1}, we can see that the spirals can be detached and attached to the planet during one orbit. Or we could think of it as that one
spiral is excited and propagates away before a new spiral is excited. Although the wake is changing { over time}, it returns to the same shape after one full orbit except that the spiral extends further radially. This periodicity of spiral shape is expected since the planet's motion has the periodicity of one planetary orbit.
For the spirals excited by an eccentric perturber, one noticeable feature is that these spirals bifurcate at some positions, which is drastically different from the spirals excited by a circular perturber. The bifurcation points, which we refer as ``V points'', are labeled in the upper left panel of Figure \ref{fig:e0p25}.

When the perturber's eccentricity increases, the spirals can change their shapes more dramatically during one orbit.  In Figures \ref{fig:e0p25} and \ref{fig:e0p5}, part of the outer spiral is inside the planet's orbit at $t=T_{orb}$. At one given radius (e.g $R=0.6 a_{pl}$ where $a_{pl}$ is the planet's semi-major axis), there are 3 spirals in Figure \ref{fig:e0p25} and even 5 spirals in Figure \ref{fig:e0p5} at $t=2T_{orb}$. All these spirals have different pitch angles and radial extent. Around the planet, the spirals cross each other. Overall, the spirals appear to be highly complex. 

The simple analytical model reproduces the spiral shapes remarkably well (the bottom panels of Figures \ref{fig:e0p1} to \ref{fig:e0p5}). Most importantly, it predicts the ``V points'' and the cross points of the spirals. The only noticeable mismatch is that the analytical model under-predicts the extend of some spirals. The spirals in simulations travel a little bit further than the ``V points''. This is similar to the discrepancy at the tip of the spirals shown in Figure \ref{fig:circular}. We will discuss this more in Section 5.

\begin{figure*}
\includegraphics[trim=10mm 0mm 5mm 0mm, clip, width=6.6in]{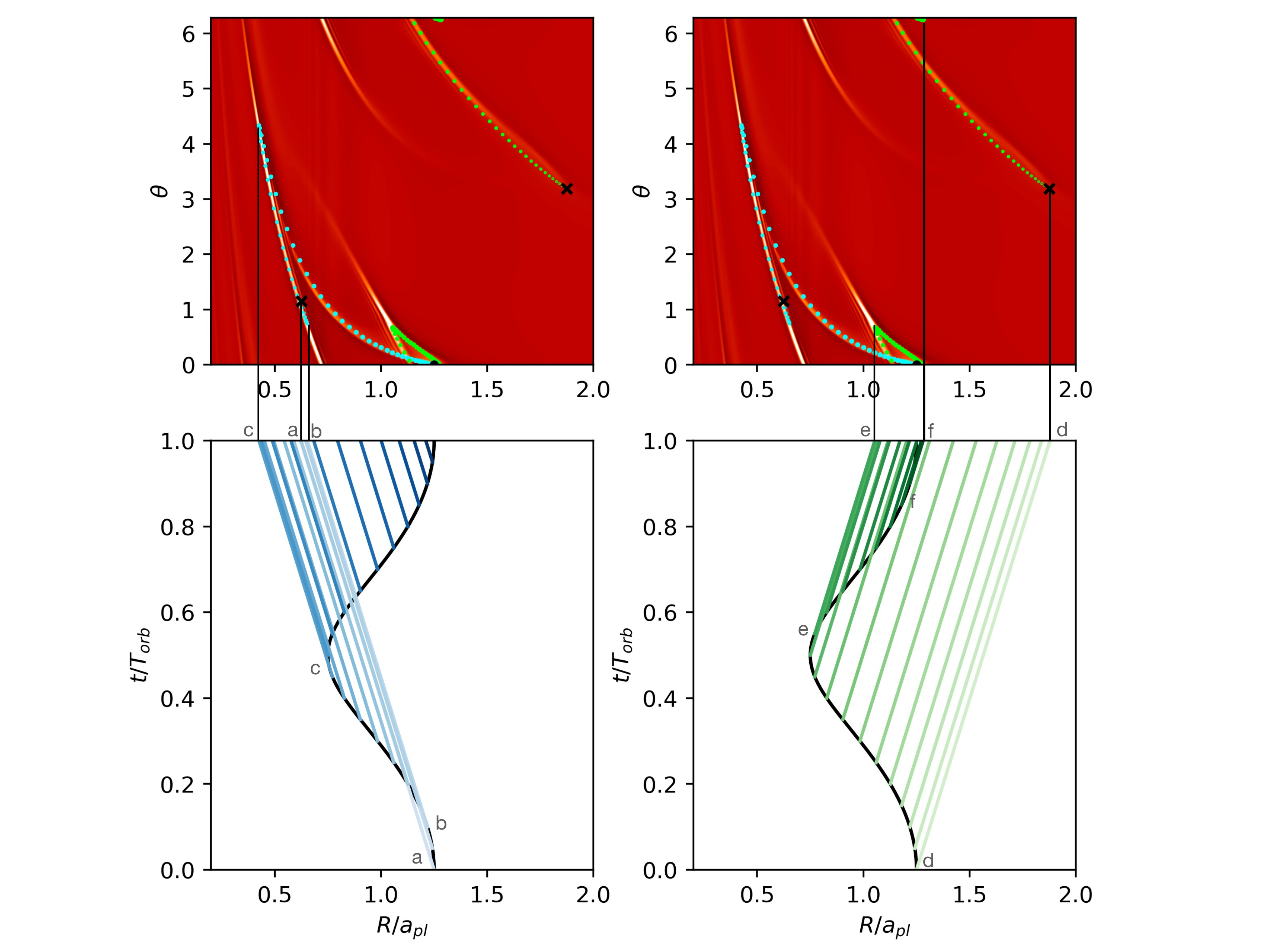}\\
\caption{The top panels: the spirals excited by an eccentric planet ($e=0.25$) at $t=T_0$, plotted in the polar coordinate system. The bottom panels: the planet's radial position with time (black curves). Each wavelet propagates inwards (the left panel) or outwards  (the right panel) following the tilted straight lines after they are excited. The wavelets' positions at $t=T_{orb}$ in the bottom panels correspond to their radial positions in the upper panels. The first wavelets that are excited at $t=0$ are labeled with the crosses in the upper panels, and correspond to the wavelets at ``a'' and ``d'' in the bottom panels. The spirals' ``V points'' (where the spirals bifurcate) correspond to the the points where the spirals make a turn, and they occur when the planet's radial speed equals the sound speed (b, c, e, f points in the bottom panels). }
\label{fig:untitled}
\end{figure*}

Successfully reproducing the ``V points'' allows us to use the analytical model to understand the origin of the spiral bifurcation. The black solid curves in the bottom panels of Figure \ref{fig:untitled} show the planet's radial position with time. Since the planet in this case has an eccentricity of 0.25, it oscillates between 1.25 $a_{pl}$ and 0.75 $a_{pl}$. During the planet's orbital motion, wavelets are emitted and traveled inwards and outwards with time. The wavelets' radial positions with time are shown by the blue and green lines in the left (for the inward moving wavelets) and right bottom panels (for the outward moving wavelets). These lines are plotted darker when the wavelets are emitted later during the orbit. The lines are straight since the sound speed is a constant in this disc. The slopes of the lines are $dt/dR=1/c_{s}(R)$.
The top panels show the spirals and the wavelets (dots) at 1 planetary orbit. These wavelets' radial positions correspond to the positions of the blue and green lines at $t=T_{orb}$ in the bottom panels. Normally, the wavelet that is emitted earlier by the planet travels further away from the planet.
But since the eccentric planet's radial motion can be faster than the sound speed occasionally, an inward moving wavelet (the left panel) that is emitted later  can be at a smaller $R$ than the wavelet that is emitted earlier. The ``V points'' occur when the planet's radial speed equals the sound speed. For a planet with eccentricity of $e$, the ``V points'' correspond to the wavelet emitted at the planet's true anomaly  $\theta$ where
\begin{equation}
\sqrt{\frac{G(M_*+M_p)}{a_{pl}(1-e^2)}}\cdot e\cdot sin\theta = \pm c_s\,, \label{eq:radialspeed}
\end{equation}
where the left side is the eccentric planet's radial speed with $\theta$.
Thus, for either the inner or outer spiral, there are two ``V points'' generated during each orbit.
The ``V points'' are labeled as `b' and 'c' for the inner spiral and `e' and `f' for the outer spiral in Figure \ref{fig:untitled}.
We can also see the generation of new ``V points'' in the bottom panels of Figure \ref{fig:e0p25}. If we just focus on the outer spiral, 2 new ``V points'' are generated from $T_{pl}$ to 2$T_{pl}$. The generation of new ``V points'' also makes the spirals more complicated with time until the spirals reach the inner and outer boundaries. { To see where the spirals originate, the first wavelets that are excited at $t=0$ are labeled with the crosses in the upper panels, and correspond to the wavelets at ``a'' and ``d'' in the bottom panels. }

\section{The Pitch Angle and Pattern Speed}
The pitch angle  ($\zeta$) and pattern speed ($\Omega_{pat}$) are two fundamental properties of the spirals. Pitch angles can be directly measured from { either density contours in simulations or observational images}. Specifically, if we plot the observational image under the polar coordinate system with $ln R$ as the x-axis, the slope of the spiral ($d\theta/d ln R$) in the image is cotangent of the pitch angle ($cot \zeta$). Pattern speeds of the spirals can be derived if we can observe the spirals multiple times with a long enough time interval. Although this is not possible for studying spiral galaxies, it is feasible for protoplanetary disc observations. { Current and future observations can probe protoplanetary discs at the AU scale where the orbital time can be less than several years}. 

\begin{figure*}
\includegraphics[trim=0mm 0mm 0mm 0mm, clip, width=6.6in]{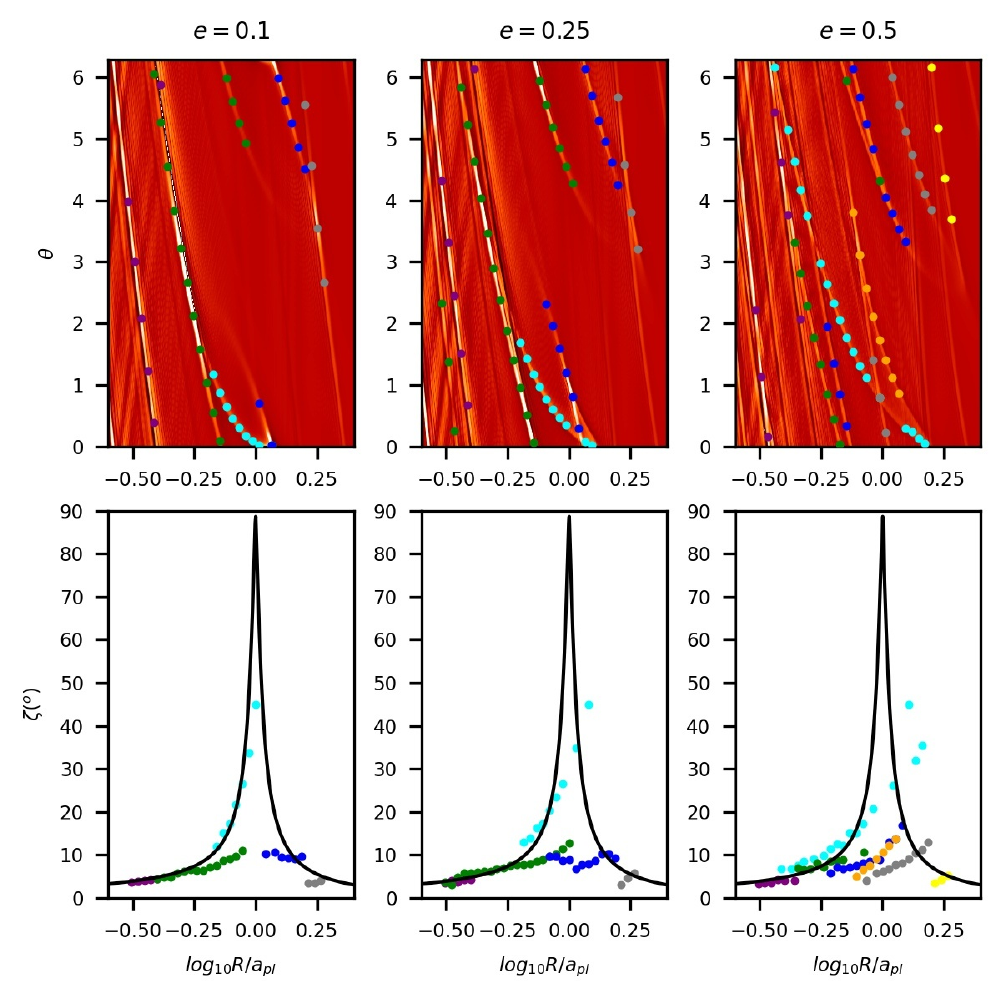}\\
\caption{The top panels: the spirals excited by eccentric planets ($e$=0.1, 0.25, 0.5, from the left to right) at $t=2 T_{orb}$. Different
spirals are labeled with different colored dots. Their pitch angles are calculated and shown in the bottom panels. The black curves in the bottom
panels are based on the analytical theory for a circular planet.}
\label{fig:pitch}
\end{figure*}

\begin{figure}
\includegraphics[trim=1mm 0mm 0mm 0mm, clip, width=3.6in]{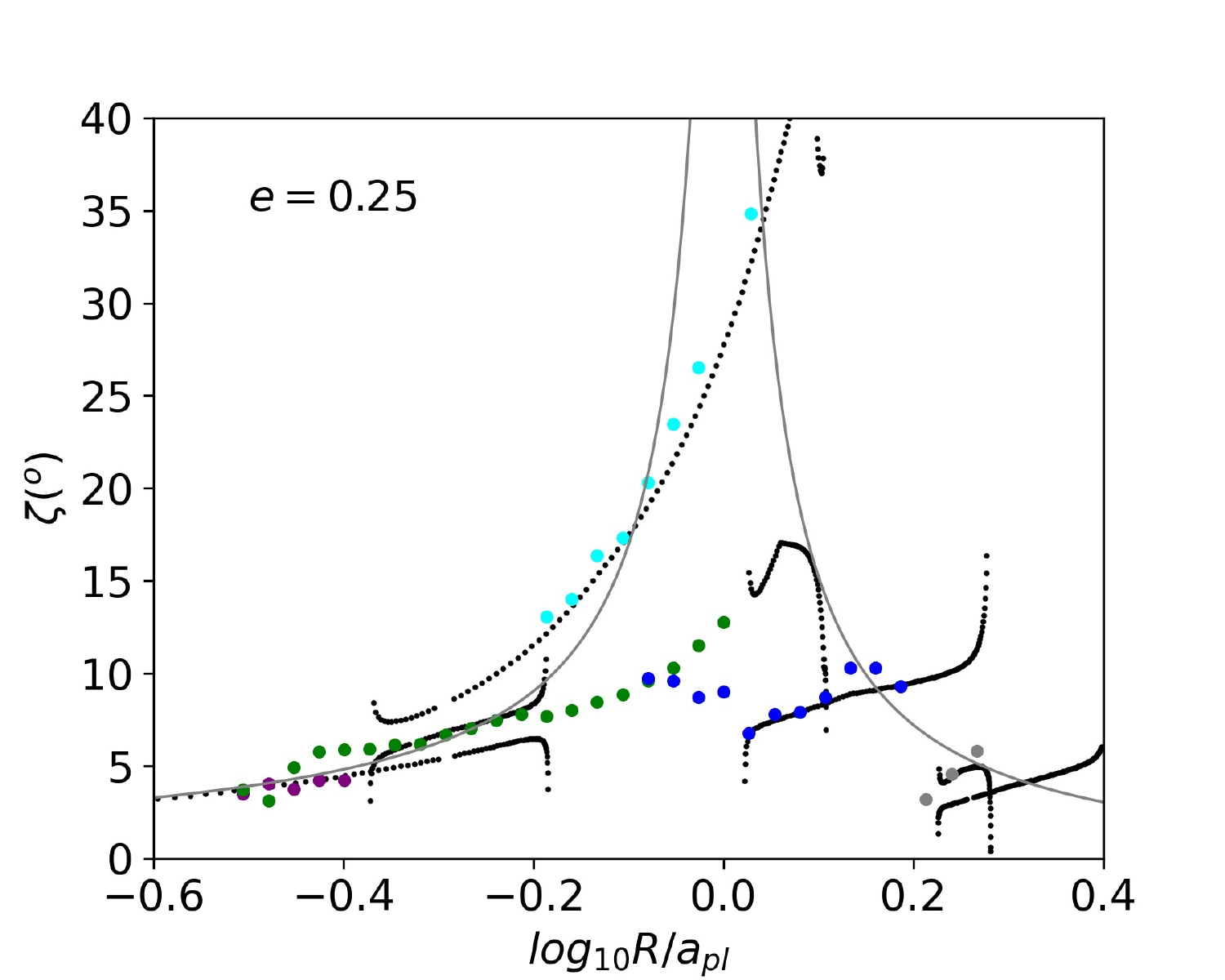}\\
\caption{ The same as the $e$=0.25 panel in Figure \ref{fig:pitch} but over-plotted with the pitch angle derived from the analytical theory (black dots).}
\label{fig:pitchana}
\end{figure}

The pitch angle of spirals excited by a circular perturber is $\sim$arccot$|(\Omega_{pat}-\Omega)R/c_s|$, shown as the black solid curves in the bottom panels of Figure \ref{fig:pitch}. To measure the pitch angle of spirals excited by the eccentric perturber, we identify density peaks on the spirals, as shown in the top panels. Different spirals are
labeled by dots with different  colors. Using the positions of these dots, we calculate the pitch angle of each spiral in the bottom panels.  
{ Most spirals (including the outer spirals) by eccentric planets have increasing pitch angles towards larger radii, which is noticeably different from the outer spirals in the circular case. A higher planet eccentricity results in more spirals and larger pitch angle deviations compared with planets on circular orbits. }
At the same radius, different spirals could have different pitch angles. The spiral with the largest pitch angle (cyan dots) is close to the planet. 

Figure \ref{fig:pitchana} shows the comparison between the measured pitch angles in the $e$=0.25 case and the pitch angles derived from our analytical solution at t=2 $T_{orb}$ (the rightmost panel in Figure \ref{fig:e0p25}). We didn't pick out all the spirals in the simulations since some spirals are very weak in Figure \ref{fig:e0p25}. Overall, we can see that the analytical model correctly predicts the pitch angles for most parts of the simulated spirals, except that the spirals in the analytical models are shorter than the spirals in simulations.  
Since the spirals by an eccentric perturber cannot maintain steady states, the pitch angle also changes with time.

\begin{figure}
\includegraphics[trim=5mm 1mm 0mm 0mm, clip, width=3.6in]{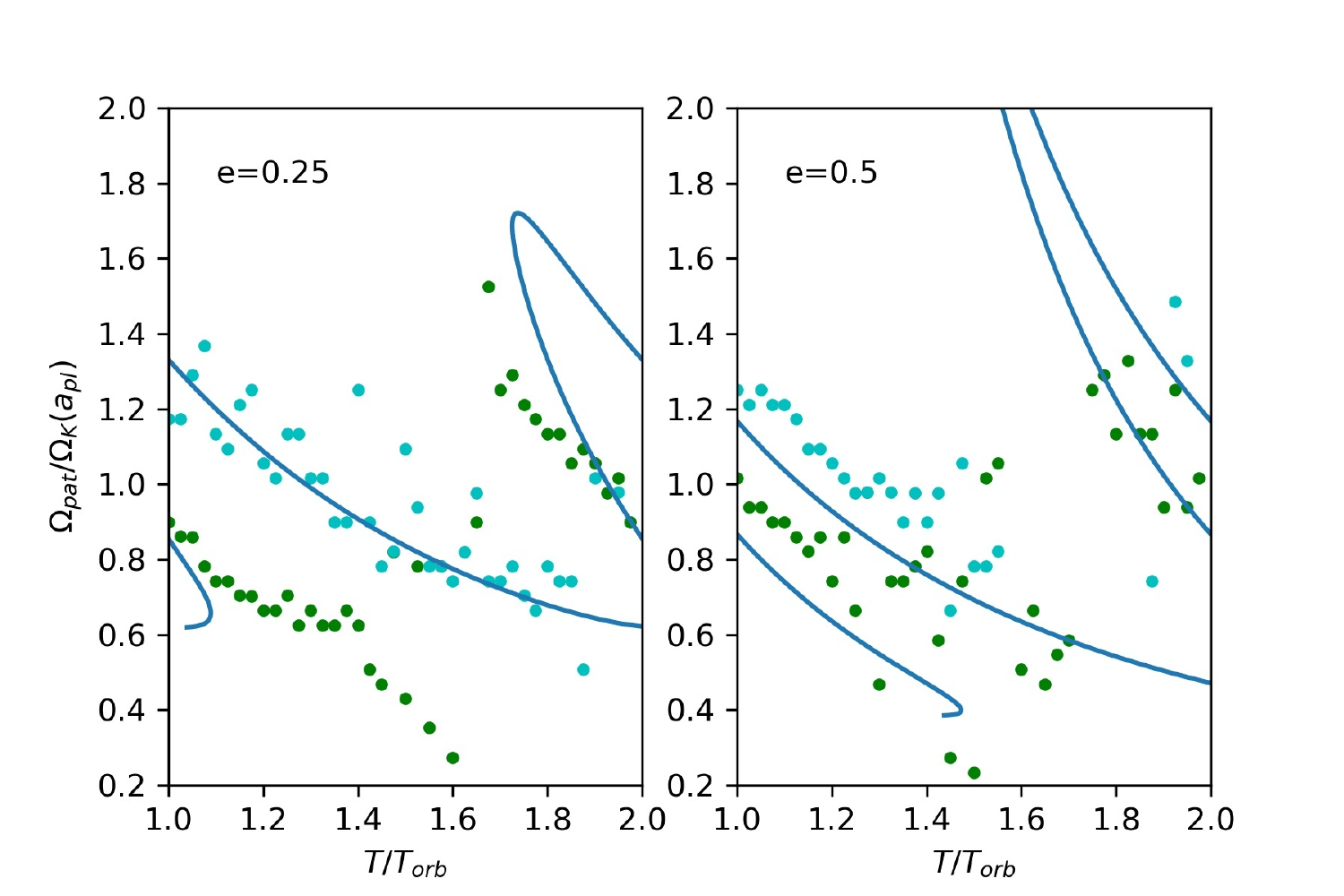}\\
\caption{The pattern speed of the spirals at  $R=0.6 a_{pl}$ from 1 to 2 $T_{orb}$. Two spirals which correspond to the cyan and green dots (the same color) in Figure \ref{fig:pitch} have been identified over the orbit. The blue curves are from the analytical theory (Equation \ref{eq:pattern}). }
\label{fig:pitchtime}
\end{figure}

To study the pattern speeds of the spirals, we identify the azimuthal position of the density peaks at $R=0.6 a_{pl}$. Two peaks associated with two spirals (cyan and green spirals in Figure \ref{fig:pitch}) have been studied. We follow these two density peaks from $T_{pl}$ to 2$T_{pl}$ to calculate the pattern speeds of the spirals at $R=0.6 a_{pl}$. The resulting pattern speeds are plotted in Figure \ref{fig:pitchtime} for e=0.25 and e=0.5 cases. In our analytical model, the spiral/wavelet at $R=0.6 a_{pl}$ should have the same pattern speed as the planet when this wavelet was emitted. Since it takes the wavelet a time of $(R_{pl}(t)-0.6 a_{pl})/c_s$ to propagate to  $0.6 a_{pl}$ after it was emitted by the planet at $t$, the spiral's pattern speed at $R=0.6 a_{pl}$ is the planet's orbital speed $\Omega_{pl}$ at a time $(R_{pl}(t)-0.6 a_{pl})/c_s$ ago. In other words, the patter speed at $R=0.6 a_{pl}$ at $t$  is
\begin{equation}
\Omega_{pat,R=0.6 a_{pl}}(t)=\Omega_{pl}\left(t-\left[R_{pl}(t)-0.6 a_{pl}\right]/c_s\right)\,.\label{eq:pattern}
\end{equation}
This pattern speed is plotted as blue curves in Figure \ref{fig:pitchtime}. These curves can also be used to estimate how many spirals exist at one radius. We can see that, during one orbit, the number  of spirals at 0.6 $a_{pl}$ can vary between 1 and 3 for the $e=0.25$ case. The analytical curves agree with the measured $\Omega_{pat}$ fairly well. Overall, the pattern speed varies with time and the two spirals have different pattern speeds. Theoretically, the range of the spiral's pattern speed is the same as the range of the planet's orbital speed, since the pattern speed equals the planet's orbital speed at some point in the past. However, the measured pattern speed  can be lower than what the theory predicts. This is because our simple theory cannot capture the spiral beyond the ``V point'' that is due to the wave dispersion (Section 5). 

\begin{figure*}
\includegraphics[trim=0mm 0.5mm 0mm 0.5mm, clip, width=5.5in]{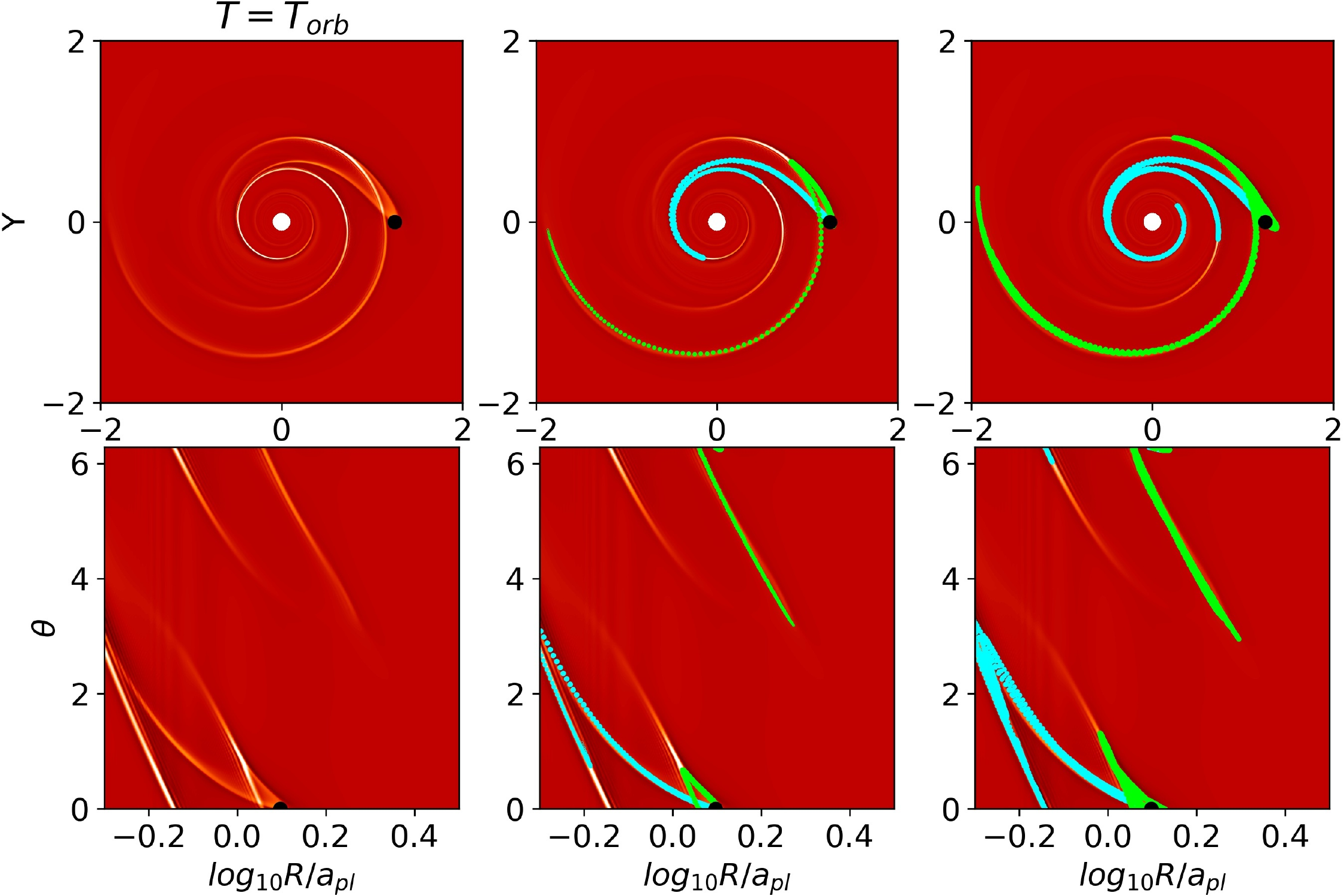}\\
\caption{The spirals excited by an eccentric planet ($e$=0.25) at $t=T_{orb}$. The middle panels are the same as those in Figure \ref{fig:e0p25}, while the right panels overplot a modified analytical model in which the wavelets have some dispersion extending over a time period of $\Delta T$= $\pm$0.15 $T_{orb}$. }
\label{fig:extendedspiral}
\end{figure*}

\section{Discussion}
\subsection{Wave Dispersion and Strength}
Despite the success of the analytical method, this simple method underpredicts the spiral's extent at the tips and ``V points'' of the spirals (middle panels in Figure \ref{fig:extendedspiral}).  We think this is because each wavelet has some intrinsic dispersion (as the dispersion in different $m$-modes discussed in the introduction) so that it is not localized as a single point. The wavelet has some finite radial and azimuthal extent. To verify this, we assume that each wavelet has some radial extent from $R_{t_1}(t_2)-0.15 c_s T_{orb}$ to $R_{t_1}(t_2)+0.15 c_s T_{orb}$ in Equations \ref{eq:rinner} and \ref{eq:router}. The resulting spirals are shown in the right panels of Figure \ref{fig:extendedspiral}. Clearly, this approach recovers the spirals beyond the tips and ``V points'' much better.  The $0.15 c_s T_{orb}$ that we add to $R_{t_1}(t_2)$ is not from rigorous theoretical calculations and it is simply chosen for a better spiral fit. 

{ This wave dispersion has also been used to explain the disappearance of the primary spiral and the appearance of the secondary spiral induced by a circularly orbiting perturber \citep{Bae2018}.  As pointed out by \cite{Miranda2019},  the formation of secondary spirals is a generic property of wave propagation regardless of the excitation mechanisms. Indeed, these secondary spirals can also be seen in all our simulations. For example, the innermost two spirals in the $e=0.1$ case of Figure \ref{fig:pitch} (the purple and green dots) originate from the same set of waves. Due to the wave dispersion, the green spiral  (the primary spiral) gradually disappears towards the inner disc while the purple spiral (the secondary spiral) starts to appear at a slightly larger $\theta$ position. When the planet mass is low, the secondary spiral is very close to the primary spiral and sometimes we treat them as a single spiral. When the planet mass is high, the secondary spiral separates from the primary spiral, forming two distinct spirals (details in \S 5.2). }

{ Besides predicting the spiral shape, our simple method also carries information on the strength of spirals through the density of wavelets.} When the wavelets are closer to each other, their contribution to the spiral wake can be coherently added so that the spiral there should be stronger. This can explain why the spirals are strong at the ``V points''. To accurately predict the strength of the spirals, a more detailed model which includes both wave excitation and wave propagation is needed.

\subsection{Spirals by High-mass Planets}
\begin{figure*}
\includegraphics[trim=0mm 13mm 0mm 10mm, clip, width=6.6in]{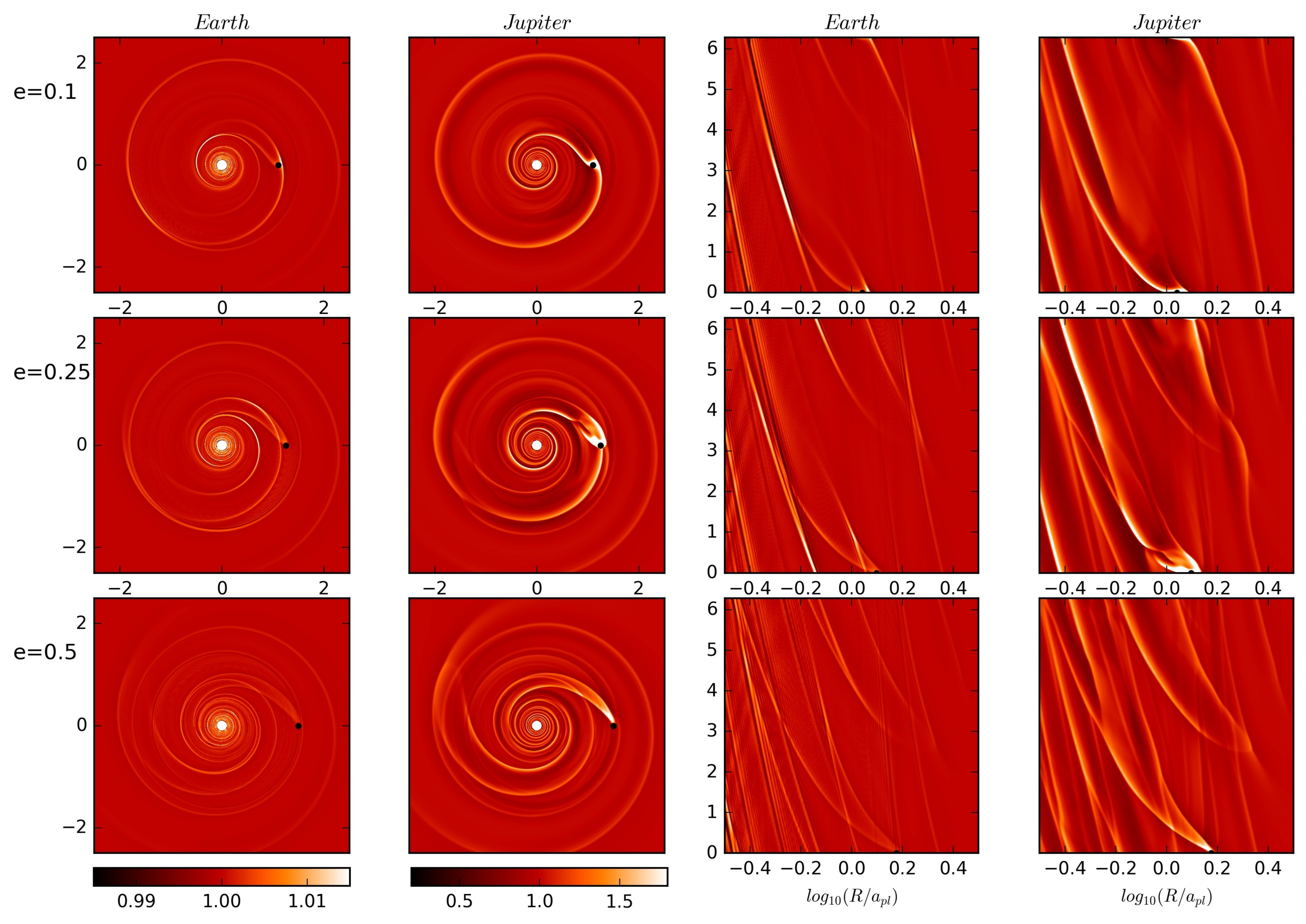}\\
\caption{{ The spirals excited by eccentric low-mass planets (one Earth mass) and high-mass planets (one Jupiter mass) at two planetary orbits. The left two columns show the density perturbation in the Cartesian coordinate system, while the right two columns show the same data in the polar coordinate system. The planet eccentricity increases from the top to bottom panels.} }
\label{fig:highermass}
\end{figure*}

\begin{figure*}
\includegraphics[trim=0mm 0mm 0mm 0mm, clip, width=6.9in]{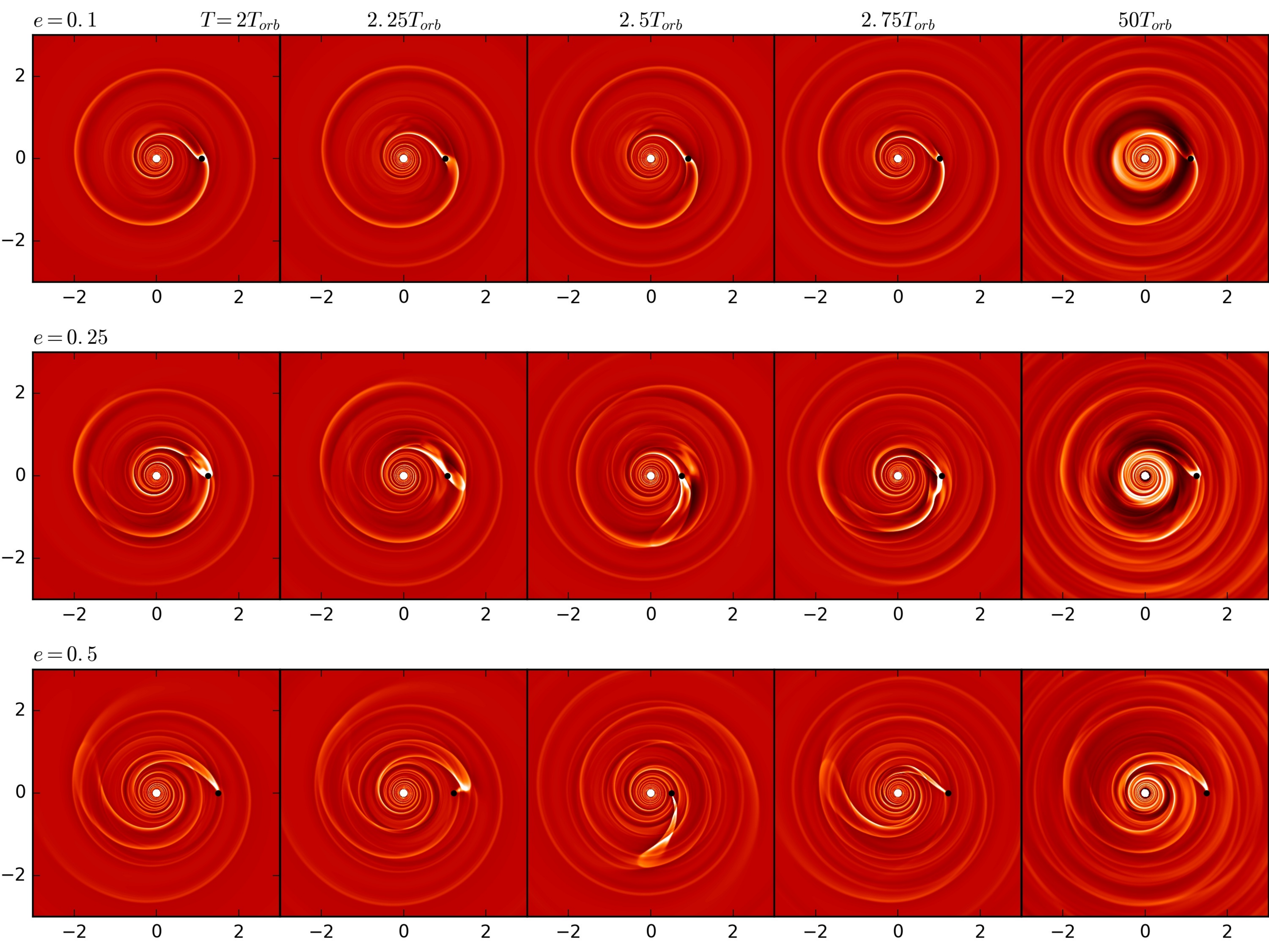}\\
\caption{{ The spirals excited by eccentric high-mass planets (one Jupiter mass) at different times (different columns). The planet eccentricity increases from the top to bottom panels.} }
\label{fig:massivelong}
\end{figure*}

{ Since our simple model is built upon the linear theory, we focus on the low-amplitude wakes excited by low-mass planets. However, spirals from low-mass planets are normally too weak to be observed \citep{Dong2015, Zhu2015c, Dong2017}. The spirals excited by a high-mass perturber are quite different from those excited by a low-mass perturber,  even if the perturber is in a circular orbit \citep{Zhu2015c}. First, the high-amplitude spirals steepen to spiral shocks \citep{Goodman2001}. Since shocks propagate faster than the sound speed, the spirals open up more with larger pitch angles than those from Equation \ref{eq:pitchangle} \citep{Bae2018b, Cimerman2021}. Second, the secondary spirals that are produced by the wave dispersion \citep{Bae2018,Miranda2019} separate from the primary spirals more \citep{Bae2018b},{ producing multiple prominent spiral arms.} Third, the spirals are more prominent at the atmosphere than at the midplane when the planet mass is higher \citep{Zhu2015c,Juhasz2018}, leading to observable spirals in near-IR observations. Due to these effects, \cite{Dong2015} suggests that a massive perturber ($\sim 10 M_{J}$) is responsible for the two near-IR spirals in MWC 758. }

{ To study the spirals excited by eccentric high-mass planets, we have carried out the same set of simulations but with a Jupiter mass planet in the disc. The resulting surface density contours are shown in Figure \ref{fig:highermass}. Although the shape of the spirals from eccentric high-mass planets are similar to those from eccentric low-mass planets, there are several noticeable differences, especially evident in the polar plots (comparing the right two columns). First, similar to the case with a circularly orbiting perturber, the spiral wakes from high-mass perturbers become quite strong and steepen into spiral shocks, and the secondary spirals start to separate  from the primary spirals. These prominent multiple spirals could be observable with near-IR observations. Second, the shape of the spirals is not as smooth as the spirals from low-mass planets. Especially when two spiral shocks cross each other (e.g., at  the spiral bifurcating or crossing points), the spirals are significantly distorted and may appear to break into several segments. The distortion is more apparent when the spiral shocks have larger amplitudes. This is expected since shock crossing is highly nonlinear and more complicated than the linear wave crossing.  }

The spiral shocks can also deposit the spirals' angular momentum to the background disc, leading to gap opening. To study the gap structure induced by a massive eccentric planet,  we continue our high-mass planet simulations to 50 planetary orbits. The discs' surface density at the end of the simulations is shown in the rightmost panels of Figure \ref{fig:massivelong}. Clearly, the gap is shallower when the planet has a higher eccentricity. This is because weaker spirals are launched by a more eccentric planet (e.g., Figure \ref{fig:highermass}). When a gap is induced (the $e=$0.1 and 0.25 cases), the gap is more dynamic when the planet's eccentricity is higher.  { Regarding the gap depth, this degeneracy between the planet mass and eccentricity may be useful for placing limits on the eccentricity of planets in gaps if the planets' mass can be independently constrained. \cite{Rabago2021} point out that the planet detection in the DSHARP sample \citep{Andrews2018} using ALMA CO kinematic observations \citep{Pinte2020} is not fully consistent with the planet mass from dust continuum observations \citep{ZhangS2018}. To be detected by CO kinematic observations, the planet's mass needs to be higher than one Jupiter mass \citep{Rabago2021,Bollati2021,Izquierdo2021}. These planets should produce very prominent dusty gaps, but most of these gaps are shallow and narrow in the dust continuum observations. One explanation is that we don't fully understand the physical processes that produce the CO kinematics or dusty gaps. Another explanation could be that the high mass planets
are in eccentric orbits, producing less prominent gaps or traveling outside the main gaps.  }

We also notice that, when the planet is at the apastron of its orbit, the planet can be at the outer gap edge. { This is similar to PDS 70c \citep{Haffert2019}, where the planet is close to the outer gap edge.} Then, the planet can potentially accrete more material from the circumstellar disc, facilitating its growth and { the circumplanetary disc formation \citep{Isella2019,Benisty2021}}.

\subsection{Applications to Observations}
{ Eccentric planets can induce a wide variety of spirals. Although this makes it easier to find a spiral that can fit
observations,  it poses a challenge to use the spirals to accurately predict the planet's mass and position. The spirals excited by eccentric high-mass planets at different times are displayed in Figure \ref{fig:massivelong}. When the planet's eccentricity is $\lesssim$ 0.1, the spirals look similar to the spirals excited by a circular planet. This is because the planet's radial speed is less than the disc's sound speed (based on Equation \ref{eq:radialspeed} with the disc's $c_s/v_K\sim$ 0.1), so that spiral bifurcation and crossing do not occur. 

When the planet's eccentricity is higher than 0.1, the spirals appear much different. First, more spirals are excited when the planet's eccentricity is higher. We caution that, when a gap is induced by a planet (which can be a circularly orbiting planet), the gap edge vortices can also generate many spirals \citep{Huang2019}. Second, the spirals can bifurcate and cross each other, making the spiral pattern complicated. The bifurcation point corresponds to the wavelet that is emitted by the planet when the planet's radial speed equals the disc's sound speed (Figure \ref{fig:untitled}). Theoretically, we can use the bifurcation point to constrain the planet's position and speed in the past, which can then be used to constrain the planet's current position. However, this may not be practical since the bifurcation point is difficult to be identified even in the simulation images without prior knowledge of the planet's orbit (Figure \ref{fig:untitled}). The bifurcation point can also be confused with the spiral crossing point. Third, probably most importantly, different spirals and even different parts of one spiral can have different pattern speeds. It is possible to use the measured pattern speeds to constrain the planet's eccentricity and current position by understanding that the pattern speed of any part of a spiral corresponds to the planet's orbital frequency in the past. This is a challenging task if we only have several pattern speed measurements. However, if the pattern speeds of more spirals or more parts of a spiral have been measured, the planet's eccentricity and position will be better constrained. Generally, the pattern speed should be between the planet's slowest and fastest orbital frequencies
\begin{equation}
\frac{\sqrt{\mu}}{a^{3/2} (1+e)}\sqrt{\frac{1-e}{1+e}}<\Omega_{pat}<\frac{\sqrt{\mu}}{a^{3/2} (1-e)}\sqrt{\frac{1+e}{1-e}}\,,\label{eq:Omegapat}
\end{equation}
where $\mu=G(M_*+M_p)$. We note that the pattern speed can actually go lower than the leftmost term in Equation \ref{eq:Omegapat} (as shown in Figure \ref{fig:pitchtime}) since our analytical model cannot capture the tip of some spirals formed by the wave dispersion.}

{ A variety of spirals have been  discovered in many protoplanetary discs \citep{Yu2019}. For example, the CO spirals in TW Hydrae \citep{Teague2019} have pitch angles dropping from 9$^o$ at 70 au to 3$^o$ at 200 au, while the spirals in HD34700A \citep{Monnier2019,Uyama2020} can reach 30$^o$ pitch angles. The small pitch angles of TW Hydrae spirals could be from the buoyancy resonances \citep{Zhu2012b} of a circularly orbiting planet \citep{Bae2021}. On the other hand, these spirals could be the outer spirals of an eccentric  perturber. As shown in the 2$T_{orb}$ panels of Figures \ref{fig:e0p1} and \ref{fig:e0p25},  the outer spiral can be considered as two separate spirals, one with a larger pitch angle joining the planet (the blue spirals in the $e=0.1$ and $e=0.25$ panels of Figure \ref{fig:pitch}) and one with a smaller pitch angle further away (the grey spirals in the same panels of Figure \ref{fig:pitch}). This is consistent with TW Hydrae  observations where several spirals seem to join together with the inner one having a larger pitch angle. 

For HD 34700A,  \cite{Monnier2019}  conclude that a circular perturber is difficult to explain those large pitch angles. Furthermore, several spirals have been detected in the disc. An eccentric perturber will naturally produce several spirals and some have large pitch angles, as in the $e=0.25$ and $e=0.5$ panels of Figure \ref{fig:pitch}. Although HD 34700A is a spectroscopic binary with $a=0.21$ au and $e=0.25$ \citep{Torres2004}, it is unlikely that such a binary can excite large scale spirals at 200 au. Another perturber (as in \citealt{Uyama2020}) may be responsible for these spirals. We note that the spiral features of HD 34700A are similar to those in HD 142527 \citep{Avenhaus2014}, which has been explained by an eccentric binary (with $a\sim30$ au) within the cavity \citep{Price2018}.

Another system with multiple spirals is AB Aurigae \citep{Boccaletti2020}. A large number of spirals are detected at near-IR \citep{Fukagawa2004,Hashimoto2011}, while two CO spirals are detected with ALMA \citep{Tang2017}. The near-IR spirals are highly complex. The spirals can bifurcate, cross, and break. They resemble the spirals excited by a moderately or highly eccentric perturber (e.g., $e=0.25$ or $e=0.5$ cases).

An eccentric perturber  can explain not only multiple irregular spirals but also some features of the grand-design two-spiral systems. MWC 758 has two prominent spirals \citep{Grady2013,Benisty2015}. These two spirals are explained with a single outer perturber \citep{Dong2015}, one inner and one outer perturber \citep{Baruteau2019}, or one eccentric inner perturber \citep{Calcino2020}. Previously, the difficulties in explaining the spirals with an inner perturber is the observed large pitch angles ($\sim$20$^o$) and the slow pattern speeds \citep{Ren2020}. With an eccentric inner perturber, its outer spirals can have large pitch angles and the pattern speeds can be low. \cite{Calcino2020} adopts an inner companion with $e=0.4$, which reproduces the two open spirals. If we use Equation \ref{eq:Omegapat} with $e=0.4$ and the $\Omega_{pat}$ measured in \cite{Boccaletti2021} and \cite{Ren2020}, we derive that the perturber's semimajor axis is still quite large $\gtrsim 170$ au, not in the cavity.  However, Equation \ref{eq:Omegapat} cannot fully capture all the spirals (Figure \ref{fig:pitchtime}) due to the wave dispersion, and an  eccentric perturber with a much smaller $a$ remains a possibility.  A bigger challenge for both the inner and outer perturber scenarios is that such a predicted massive companion has not been discovered yet \citep{Boccaletti2021}.

Another grand-design spiral system SAO 206462 \citep{Muto2012} exhibits different pattern speeds for its two spirals \citep{Xie2021}. Although two planets (one at 120 au and one at 49 au) have thus been suggested \citep{Xie2021}, an eccentric perturber may also explain these two independent-moving spirals. If we use Equation \ref{eq:Omegapat}, we can derive that the ratio between the minimum and maximum pattern speeds is $(1-e)^2/(1+e)^2$. Thus, if both spirals are induced by one eccentric perturber, we have 
\begin{equation}
\left(\frac{1-e}{1+e}\right)^2\leq\left(\frac{49}{120}\right)^{1.5}\,,
\end{equation}
or $e\geq$0.32. If we adopt $e=0.4$, the planet's semi-major axis should be between 74 and 92 au. However, many more spirals should be excited by such an eccentric perturber, which needs to be tested by future observations. { Overall, considering the complex spiral pattern excited by eccentric perturbers, studying pattern speed of spirals could be an important way to locate the planet.} 
} 

\section{Conclusion}
{ We develop a general method to calculate the shape of spirals launched by perturbations in a disc (e.g., from disc turbulence, vortices, or gravitational perturbers). We apply this method to spirals launched by an eccentric planet, and find good agreement with numerical simulations.  
The spirals excited by an eccentric perturber are very different from the spirals excited by a circular perturber. The spirals can detach from the perturber, bifurcate, or even cross each other. Multiple spirals can be excited. Different spirals or even different parts of one spiral can have different pitch angles and pattern speeds. These pitch angles and pattern speeds are also changing over time. 

With a high-mass eccentric planet, the spirals steepen to shocks. Secondary spirals start to separate from the primary spirals, forming more spirals. The crossing of spiral shocks lead to distorted or broken spirals. Gaps can be induced by high-mass eccentric planets. However, the gap induced by a more eccentric planet is shallower due to the weaker spiral shocks it launches. The eccentric planet can travel from one gap edge to another during its eccentric orbit,  { potentially accrete more material from the circumstellar disc, and facilitate the planet's growth and the circumplanetary disc formation.}

Finally, we discuss the potential applications of the theory. The multiple spirals in TW Hydrae, HD34700A, AB Aurigae and the different pattern speeds of spirals in HD 135344 could be indicators of eccentric perturbers in discs. Our simple method provides an easy way to calculate the spirals from an eccentric perturber, which enables using spirals to probe eccentric planets in protoplanetary discs and constrain planet formation theory. }

\section*{DATA AVAILABILITY}
The data underlying this article are available in the article and in its online supplementary material.

\section*{Acknowledgments}
We thank the reviewer for a thorough and helpful report.
Z. Z. acknowledges support from the National Science Foundation under CAREER Grant Number AST-1753168. 

\bibliographystyle{mnras}
\input{msMNRAS.bbl}

\bsp
\label{lastpage}
\end{document}